\documentclass[11pt,a4paper]{article}
\usepackage{url}
\usepackage{jcappub}
\usepackage{amsfonts}
\usepackage{subfigure}
\usepackage{graphicx}
\usepackage{ulem}
\usepackage{appendix}

\newcommand{\be}{\begin{equation}}
\newcommand{\ee}{\end{equation}}
\newcommand{\bea}{\begin{eqnarray}}
\newcommand{\eea}{\end{eqnarray}}

\newcommand{\beq}{\begin{equation}}
\newcommand{\eeq}{\end{equation}}

\begin{document}

%%%%%%%%%%%%%%%%%%%%%%%%%%%%%%%%%%%%%%%%%%%%%%%%%%%%%%%%%%%%%%%%%%%%%%
% Frontpage %%%%%%%%%%%%%%%%%%%%%%%%%%%%%%%%%%%%%%%%%%%%%%%%%%%%%%%%%%
%%%%%%%%%%%%%%%%%%%%%%%%%%%%%%%%%%%%%%%%%%%%%%%%%%%%%%%%%%%%%%%%%%%%%%

%\subheader{\hfill}

\title{Heavy axion-like particles and core-collapse supernovae: constraints and impact on the explosion mechanism}

\author[a]{Giuseppe Lucente,}
\author[a,b]{Pierluca Carenza,}
\author[c]{Tobias Fischer,}
\author[d]{Maurizio Giannotti,}
\author[a,b]{Alessandro Mirizzi}

\affiliation[a]{Dipartimento Interateneo di Fisica ``Michelangelo Merlin'', Via Amendola 173, 70126 Bari, Italy}
\affiliation[b]{Istituto Nazionale di Fisica Nucleare - Sezione di Bari, Via Orabona 4, 70126 Bari, Italy}
\affiliation[c]{Institute for Theoretical Physics, University of Wroc\l{}aw, Pl. M. Borna 9, 50-204 Wroc\l{}aw, Poland}
\affiliation[d]{Physical Sciences, Barry University, 11300 NE 2nd Ave., Miami Shores, FL 33161, USA}

\emailAdd{g.lucente5@studenti.uniba.it, pierluca.carenza@ba.infn.it, tobias.fischer@ift.uni.wroc.pl, MGiannotti@barry.edu, alessandro.mirizzi@ba.infn.it}

%\affiliation[a]{Dipartimento Interateneo di Fisica ``Michelangelo Merlin'', Via Amendola 173, 70126 Bari, Italy}
%\affiliation[b]{Istituto Nazionale di Fisica Nucleare - Sezione di Bari, Via Amendola 173, 70126 Bari, Italy}
%\emailAdd{alessandro.mirizzi@ba.infn.it}

\abstract{Heavy axion-like particles (ALPs), with masses $m_a \gtrsim 100$~keV, coupled with photons, would be copiously produced in a supernova (SN) core
via Primakoff process and photon coalescence. Using a state-of-the-art SN model, we revisit  the energy-loss SN 1987A bounds on axion-photon coupling. Moreover, we point out that heavy ALPs with masses $m_a \gtrsim 100$~MeV and axion-photon coupling $g_{a\gamma} \gtrsim 4 \times
10^{-9}$~GeV$^{-1}$ would decay into photons behind the shock-wave producing a possible enhancement in the  energy deposition that would boost the SN shock revival.
 }
\maketitle

%%%%%%%%%%%%%%%%%%%%%%%%%%%%%%%%%%%%%%%%%%%%%%%%%%%%%%%%%%%%%%%%%%%%%%%%%%%%%%%%%%%%%%%%%
\section{Introduction}

Axion-like-particles (ALPs) with masses $m_a$ in the  keV-MeV range  emerge in different extension of the Standard Model, 
as Pseudo-Goldstone bosons of some broken global symmetry (see e.g. Sec.~6.7 of Ref.~\cite{DiLuzio:2020wdo} for a recent review).
Besides QCD axions, heavy ALPs emerge in compactification scenarios of  string theory~\cite{Svrcek:2006yi,Arvanitaki:2009fg,Cicoli:2012sz}, or in the context of  ``relaxion''
models~\cite{Graham:2015cka}.
Heavy ALPs have also recently received considerable attention in the context of Dark Matter model-building. 
Indeed, they may act as mediators for the interactions between
the  Dark Sector and Standard Model (SM) allowing to reproduce the correct Dark Matter relic abundance via thermal freeze-out~\cite{Boehm:2014hva,Hochberg:2018rjs}. 
ALPs with masses below the MeV scale can have a wide range of implications for cosmology and astrophysics (see~\cite{Dolan:2017osp}
for a review), affecting for example the Big Bang Nucleosynthesis (BBN), 
the Cosmic Microwave Background (CMB)~\cite{Cadamuro:2010cz,Cadamuro:2011fd,Depta:2020wmr} and the evolution of stars~\cite{Carenza:2020zil}. 
Colliders and beam-dump experiments are also capable to explore this mass range, 
indeed reaching the $m_a \sim \mathcal{O}$(GeV) frontier, which is not covered by 
any astrophysical or cosmological considerations~\cite{Dolan:2017osp,Jaeckel:2015jla,Dobrich:2019dxc,Banerjee:2020fue}. 

Core-collapse supernovae (SNe) represent a valuable cosmic laboratory to probe ALPs \cite{Brockway:1996yr,Grifols:1996id,Payez:2014xsa}. 
In this minimal scenario, in which ALPs couple only with photons with an effective two-photons vertex
$g_{a\gamma}$, their dominant emission process in SNe is constituted by the Primakoff process on free protons,
 $\gamma + p \to p +a$, i.e. the conversion of a photon into an 
ALP in the electric field of protons in the stellar matter.
Moreover,  in a medium of sufficiently high density, two photons can annihilate producing an axion, in the so called ``photon coalescence'' or ``inverse decay process''. This effect has a kinematic threshold, vanishing for $m_a<2 \omega_{\rm pl}$, where the plasma frequency $\omega_{\rm pl}$ is the ``effective photon mass''. This process, typically neglected in previous studies,  in a SN core starts to be important for $m_a \gtrsim 10$~MeV.
The ALP emission from SNe has been used to obtain constraints on the photon-ALP coupling $g_{a\gamma}$ from 
the SN 1987A neutrino burst~\cite{Masso:1995tw,Dolan:2017osp,Lee:2018lcj}.
Indeed, for values of the ALP-photon coupling $10^{-9}$~GeV$^{-1} \lesssim g_{a\gamma} \lesssim 10^{-5}$~GeV$^{-1}$, ALPs would have contributed to an excessive \textit{energy-loss} in the SN core,
shortening the observed neutrino burst.
%Indeed for values of $g_{a\gamma} \gtrsim 10^{-9}$~GeV$^{-1}$, ALPs would have contributed to an excessive \emph{energy-loss} in the SN core,
%shortening the observed neutrino burst. 
%The couplings  $g_{a\gamma} \lesssim 10^{-5}$~GeV$^{-1}$ have also been excluded. 
%Indeed, in this range  ALPs would be trapped in 
%a SN core, contributing to an excessive \emph{energy-transfer}. %\red{[G.L: Should we explicitly talk about the ``modified luminosity criterion'' also here in the Introduction?]} 
Furthermore, in~\cite{Giannotti:2010ty,Jaeckel:2017tud} it has been shown that for coupling  $g_{a\gamma} < 10^{-9}$~GeV$^{-1}$, in the mass range $m_a \in [1, 100]$~MeV
a further constrain can be obtained from the non-observation of a gamma-ray flux from decaying ALPs, in coincidence with the SN 1987A neutrino burst.

The goal of our paper is to take a fresh look to the SN 1987A energy-loss argument on heavy ALPs, characterizing the ALP emissivity using the state-of-the-art SN simulations. Moreover, we will examine the possible impact of ALPs on the SN explosion mechanism, in scenarios in which these particles decay into photons  behind the SN shock-wave, helping its revitalization. The plan of our work is as follows. In Sec.~\ref{sec:simul} we present our SN reference model. In Sec.~\ref{sec:emiss} we characterize the ALP emissivity from Primakoff and photon coalescence processes. Sec.~\ref{sec:bound} presents our update of the  bounds on heavy ALPs from SN 1987A. In Sec.~\ref{sec:endep} we discuss the possible impact of decaying ALPs behind  the shock-wave on the SN explosion mechanism. Finally, in Sec.~\ref{sec:concl} we summarize our results and we conclude. Two Appendices follow. In Appendix~\ref{app:trapmethods} we compare the different constraining criteria in the trapping regime, while in Appendix~\ref{app:uncertainties} we show how a change in the SN progenitor mass affects the bound.

 \section{Input of our calculation}
 \label{sec:simul}
 \subsection{SN reference model}
 \label{subsec:SNmodel}

In this  work, we consider as SN reference model {\tt AGILE-BOLTZTRAN}, which is based on spherically symmetric neutrino-radiation hydrodynamics with accurate three-flavor Boltzmann neutrino transport~\cite{Mezzacappa:1993gn,Liebendoerfer:2002xn}, including a complete set of standard weak interactions (see Table~1 in Ref.~\cite{Fischer:2018kdt}). The SN simulations are launched from the 18~M$_\odot$ progenitor from the stellar evolution calculations of Ref.~\cite{Woosley:2002zz}.

Neutrino-driven explosions cannot be obtained in spherically symmetric simulations, except for low-mass progenitor stars with masses of $\text{M}\simeq 8-9$ M$_\odot$ (featuring O-Ne-Mg cores~\cite{Nomoto:1987,Jones:2013} and being associated with electron-capture SN~\cite{Huedepohl:2009wh}). Thus  enhanced neutrino heating rates have been applied here in order to trigger the SN explosion onset at a post-bounce time $t_{\text{pb}}\approx220\,\text{ms}$, following the procedure developed in Ref.~\cite{Fischer:2009af}. Once the explosion proceeds, the standard rates are restored. This artificial tool does not affect our results since the ALP production becomes relevant on a timescale on the order of one second after the supernova explosion has been launched. As a matter of fact, the supernova evolution at $t\gtrsim \mathcal{O}(1~\text{s})$ is moderately independent from the details of the explosion mechanism and it can be well simulated in a spherical symmetry since multidimensional phenomena play a minor role for determining the structure at the proto-neutron star (PNS) interior where our focus will be on the high-density and high-temperature domain. 

%%%%%%%%%%%%%%%%%%
\begin{figure} [t!]
 	\centering
 		\includegraphics[scale=0.17]{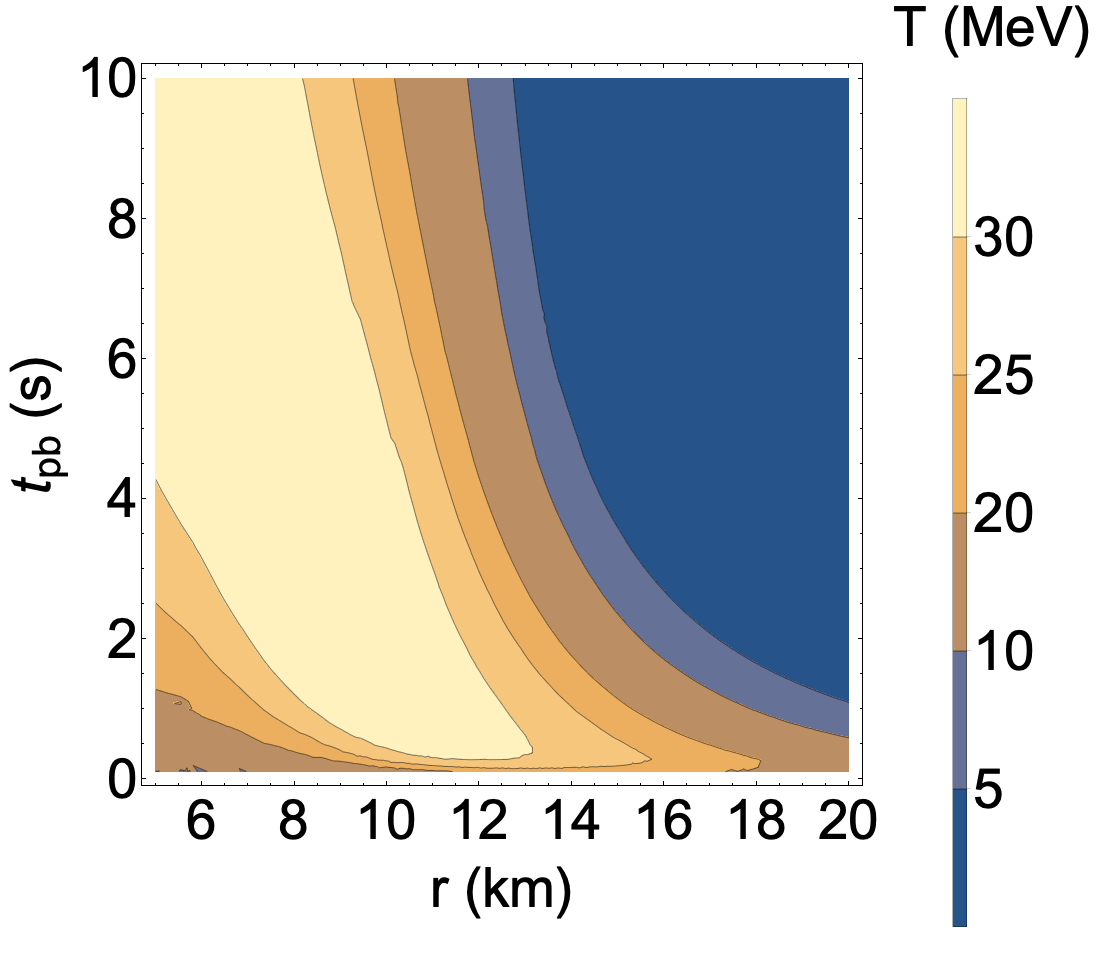} 
 		\includegraphics[scale=0.17]{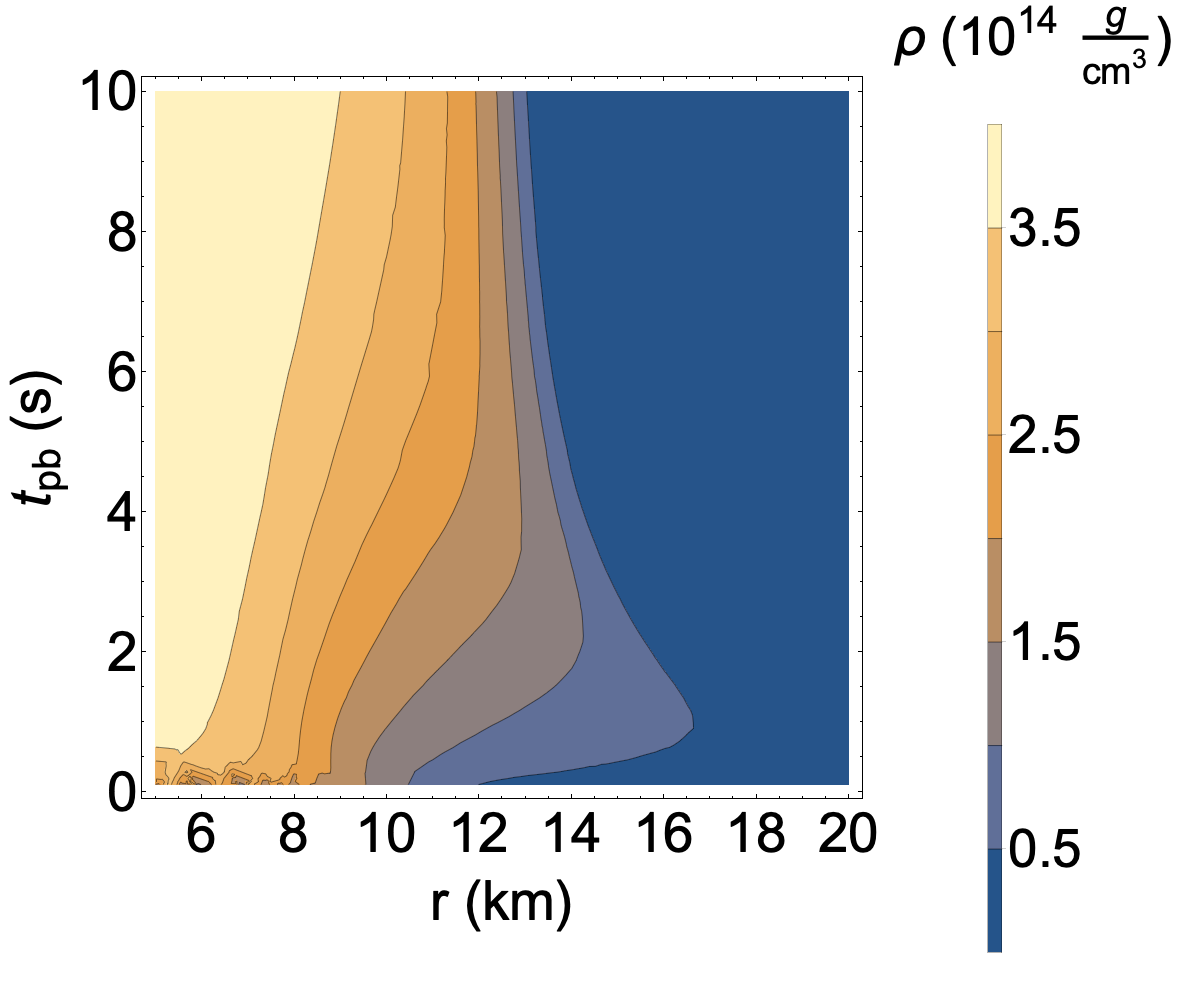}\\
 		\vspace{2pt}
 		\includegraphics[scale=0.17]{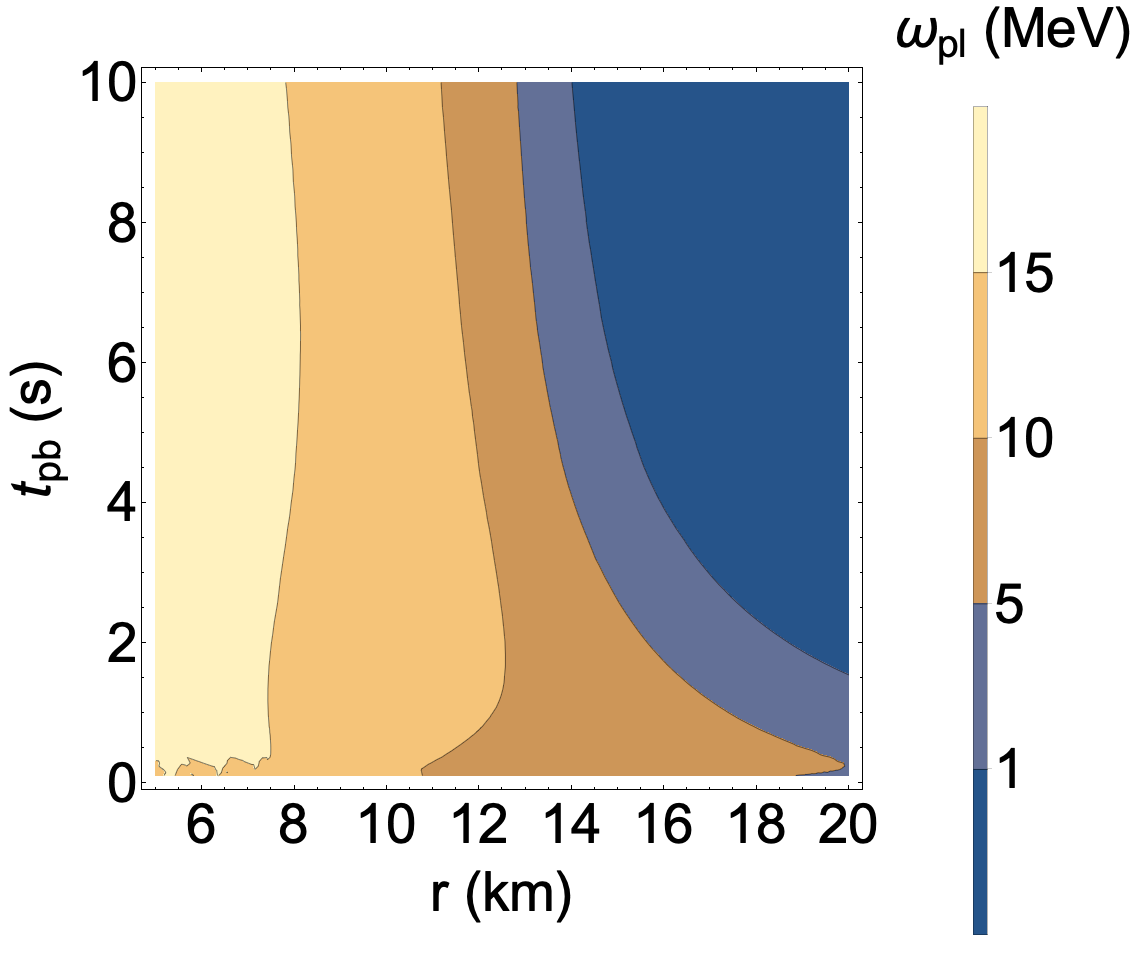}
 	\caption{Isocontours of the temperature $T$ (upper left panel), of the matter density $\rho$ (upper right panel) and of the plasma frequency $\omega_{\rm{pl}}$ (lower panel) in the plane $r$--$t_{\text{pb}}$. }
 	\label{Tempev}
 \end{figure} 
 %%%%%%%%%%%%%

%For the calculation of the ALP emissivity, we are interested in the deepest SN regions. In particular, in Fig.~\ref{Tempev} we show in the plane of radial coordinate $r$ vs post-bounce time $t_{\rm pb}$ the isocontours of temperature $T$ (left panel) and density $\rho$ (right panel). Due to the proto-neutron star  contraction, the core density monotonically increases till it exceeds the nuclear saturation density { ($\rho_{\rm sat}\approx 2.5\times 10^{14}$ g cm$^{-3}$)} for $r \lesssim 10$~km. On the other hand, at early times after the core bounce ($t_{\text{pb}}\sim 1-3\,\text{s}$) the highest temperatures [$T\sim O(40\,\text{MeV})$] are reached at $r \approx 6-12$~km. At late times ($t_{\text{pb}}\sim 10$~s), the temperature is the highest at the centre and decreases with the radius. 
  For the calculation of the ALP emissivity, we are interested in the deepest SN regions. In particular, in Fig.~\ref{Tempev} we show in the plane of radial coordinate $r$ vs post-bounce time $t_{\rm pb}$ the isocontours of the temperature $T$ (upper left panel) and the density $\rho$ (upper right panel). At early times after the core bounce ($t_{\text{pb}}\sim 1-3\,\text{s}$), the highest temperatures [$T\sim O(40\,\text{MeV})$] are reached at $r \approx 6-12$~km, while at later times ($t_{\text{pb}}\sim 10$~s), the temperature is the highest at the centre and decreases at larger distances. On the other hand, due to the proto-neutron star  contraction, the core density monotonically increases till it exceeds the nuclear saturation density { ($\rho_{\rm sat}\approx 2.5\times 10^{14}$ g cm$^{-3}$)} for $r \lesssim 10$~km. A quantity strictly dependent on the matter density $\rho$ is the plasma frequency $\omega_{\rm pl}$, playing the role of an ``effective photon mass'', which is an important factor in computing the photon coalescence rate. In a SN core $\omega_{\rm pl}\simeq 16.3~\text{MeV}~Y_e^{1/3}~\rho_{14}^{1/3}$ \cite{Kopf:1997mv}, where  $\rho_{14}=\rho/10^{14}~\text{g cm}^{-3}$ and $Y_e$ is the electron fraction. As shown in the lower panel of Fig.~\ref{Tempev}, at a fixed time, the plasma frequency is maximal in the inner SN core and monotonically decreases with the radius. In particular, at $t_{\rm{pb}}=1$~s, $\omega_{\rm{pl}}\approx15$ MeV in the interior of the proto-neutron star and it becomes smaller than 1 MeV at radii $r\gtrsim20$~km.

%%%%%%%%%%%%%%%
\subsection{Effective proton mass and chemical potential}
%%%%%%%%%%%%%%%%%%%%%%%%

\begin{figure} [t]
\centering
\includegraphics[scale=0.17]{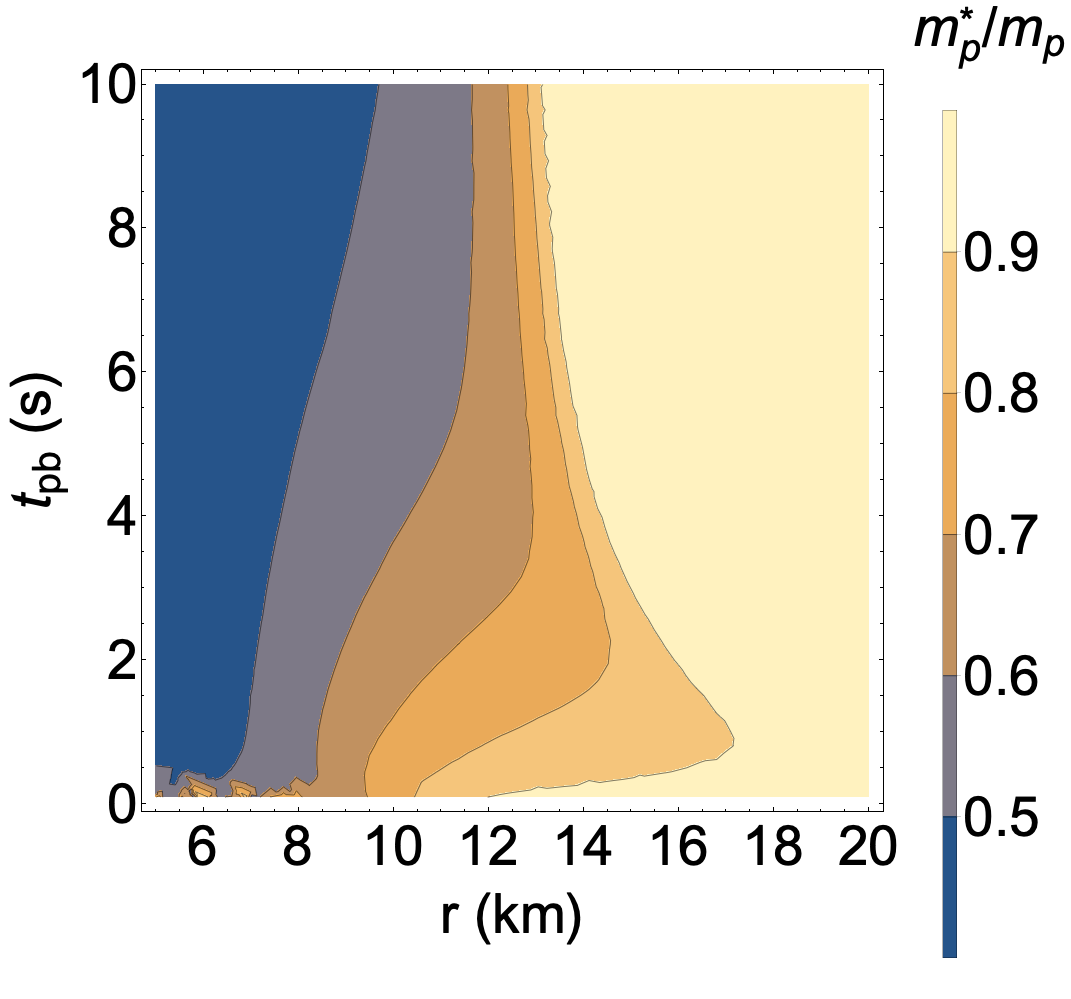}   
\includegraphics[scale=0.17]{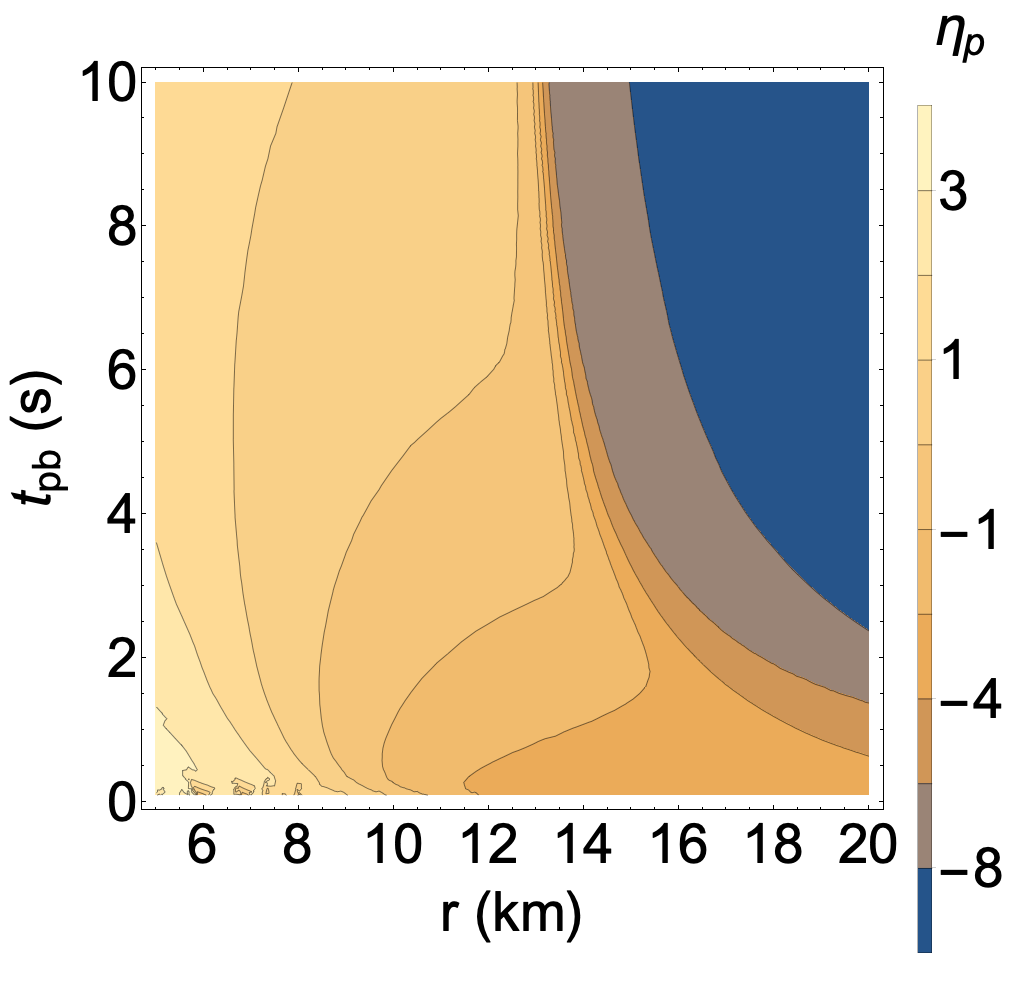}\\
\caption{Isocontours of the ratio $m_p^*/m_p$ (left panel) and of proton degeneracy parameter $\eta_p$ (right panel) in the plane $r$--$t_{\text{pb}}$.}
\label{meffetacontour}
\end{figure}

In order to evaluate the ALP production rate, in particular the axion emission due to the Primakoff process, we have to consider two nuclear matter aspects, accounted in SN simulations: the reduction of the nuclear masses due to medium effects and the possible degeneracy of protons. Both phenomena depend on the nuclear equation of state for which the relativistic mean field model of Ref.~\cite{Hempel:2009mc} is employed here (further details can be found in Ref.~\cite{Hempel:2011mk,Hempel:2014ssa}). In particular, in the hot and dense supernova matter protons are in chemical and thermal equilibrium and obey the Fermi-Dirac statistics with the following distribution function $f_p$,
\begin{equation}
f_p\left(p;\{\mu_p,T\}\right) = \left[\exp\{\beta(E(p)-\mu_p^*)\}+1\right]^{-1} \,\ ,
\label{eq:fddistrib}
\end{equation}
with inverse temperature $\beta=1/T$ and effective chemical potential $\mu_p^*$. Commonly used modern nuclear equations of state for supernova studies consider the strongly interacting nucleons at the mean field level~\cite{Hempel:2009mc,Hempel:2011mk}, which are based on the single-particle self energy, $\Sigma$, that can be separated into scalar ($S$) and the vector parts ($V$). This leads to the nucleon energy dispersion relation, $E(p)=\sqrt{p^2 + m^*_p}$
%\red{+}\Sigma_V$ \tf{[{\bf TF: check the sign in front of $\Sigma_V$}}\red{ I followed Eq. 6 in Ref. \cite{Hempel:2014ssa} in which the sign is +]},
with the proton effective mass, $m^*_p=m_p-\Sigma_p^S$, and the definition of the effective chemical potential as follows, $\mu_p^*=\mu_p-\Sigma_p^V$, related to the proton thermodynamic chemical potential, $\mu_p$. Note that we distinguish proton from neutron self energies here, in accordance with Ref.~\cite{Hempel:2011mk}. The difference between neutron and proton self energies is related with the nuclear symmetry energy, which has a strong density dependence~\cite{Fischer:2014}. In the left panel of Fig.~\ref{meffetacontour} the isocontours of {the ratio $m_p^*/m_p$} in the $r-t_{\text{pb}}$ plane are shown. We observe that at $r< 10\text{ km}$ the effective proton mass is reduced by $50\%$ with respect to its vacuum value. Indeed, the effective mass reduces steeply with increasing density above the nuclear saturation density, due to the strong density dependence of the scalar interactions in relativistic mean-field nuclear equations of state. 
  
%%%%%%%%%%%%%%%%%%%%%%%%
%%%%%%%%%%%%%%%%%%%%%%%%
  
Note further that in our calculations of the axion emission rate, we assume non-relativistic protons and the argument in the exponent of the Fermi-Dirac distribution function becomes \cite{Hempel:2014ssa},
\begin{equation}
E(p)-\mu_p^* \approx m_p + \frac{p^2}{2 m^*_p} + U_p - \mu_p \,\ ,
\end{equation}
with the definition of the proton mean-field potential, $U_p=\Sigma_p^V-\Sigma_p^S$, that allows us to rewrite the Fermi-Dirac distribution as follows,
\begin{equation}
f_p\left(p;\{\mu_p,T\}\right) \approx \left[\exp\left\{\frac{p^2}{2 m^*_p T}-\eta_{p}\right\}+1\right]^{-1}\,\ ,
\label{eq:fermidirac}
\end{equation}
where we have introduced the proton degeneracy parameter, $\eta_p$, as follows,
\begin{equation}
{\eta}_p = \frac{\mu_p -m_p - U_p}{T}\,\ .
\label{eq:eta}
\end{equation}
The right panel of Fig.~\ref{meffetacontour} shows that protons are partly degenerate ($\eta_p>1$) in the inner core of the proto-neutron star ($r< 10\text{ km}$), where the nuclear saturation density $\rho_{\rm sat}$ is reached.

\begin{figure} [t]
	\centering
	\includegraphics[scale=0.17]{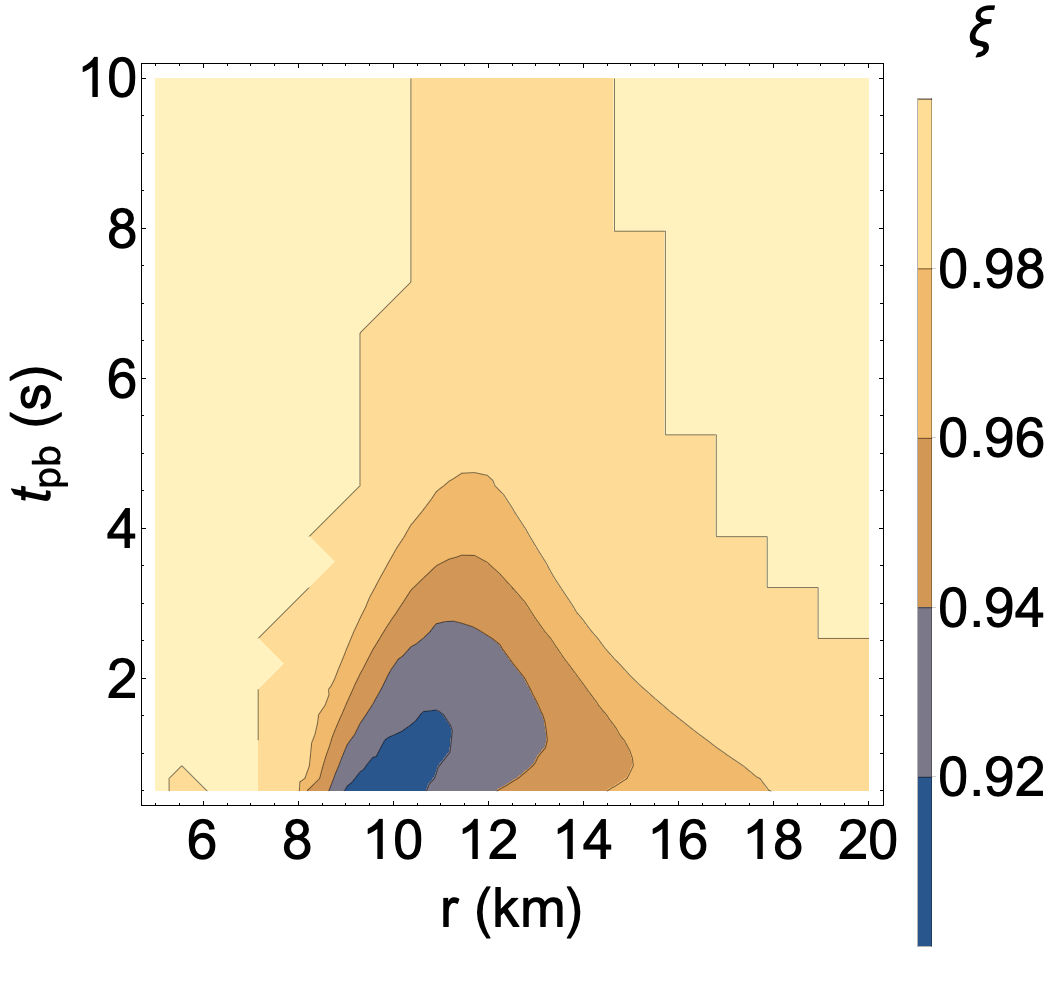} 
	\includegraphics[scale=0.17]{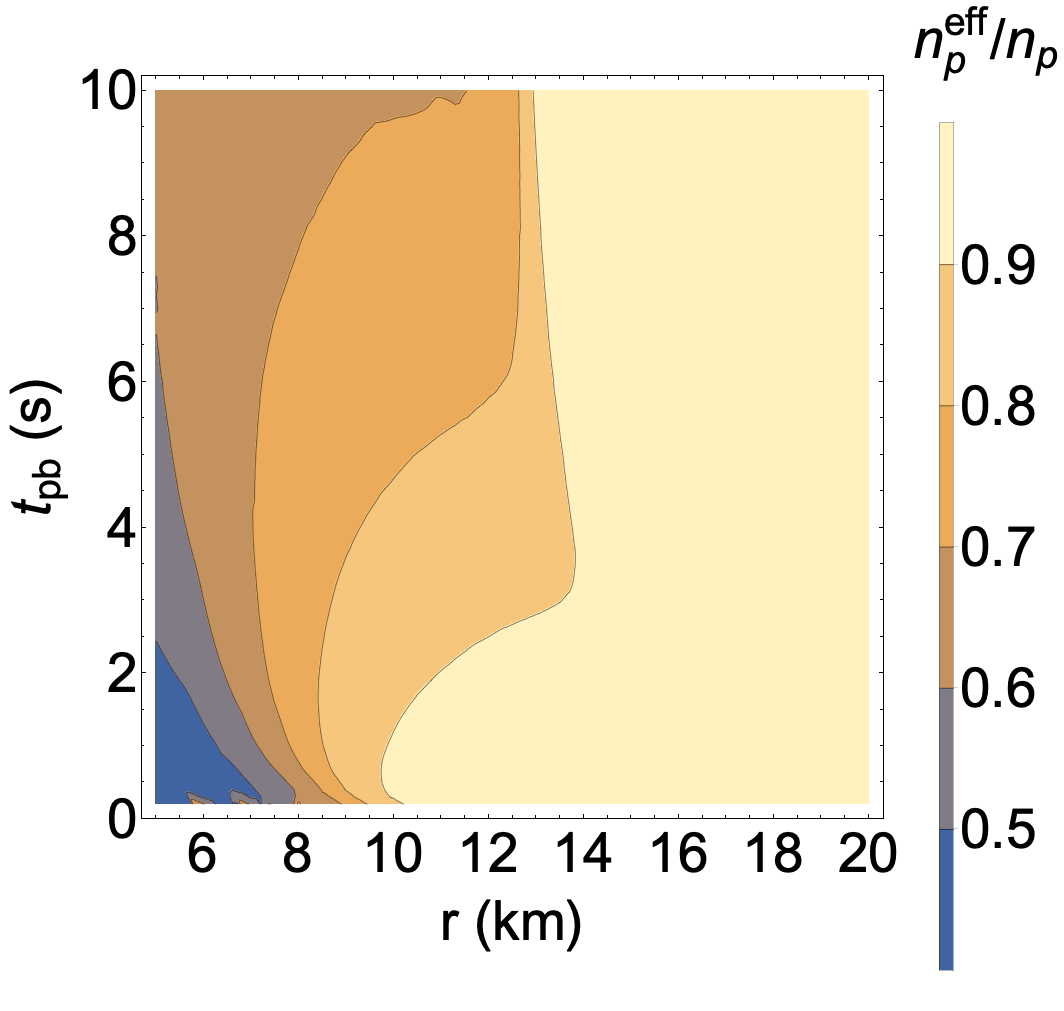}
	\caption{Isocontours of the filling factor $\xi$ (left panel) and  of the ratio $n_p^\text{eff}/n_p$ (right panel) in the plane  $r$--$t_{\text{pb}}$.}
	\label{xineffcontour}
\end{figure}

Through the distribution function of Eq.~(\ref{eq:fermidirac}), the free-proton number density can be obtained as
%%%%%%%%%
\begin{equation}
n_p=2\,\xi\int\dfrac{d^3\textbf{p}}{\left(2\pi\right)^3}f_p\,,
\label{np}
\end{equation}
%%%%%%%%%%%%%%%%%
where $2$ is the spin degeneracy factor while $\xi<1$ is the so-called ``filling factor'', which refers to unbound nuclei. It is related with the excluded volume approach of Ref.~\cite{Hempel:2009mc} employed in the nuclear equation of state to account for the dissolving of nuclear clusters with increasing density, i.e. the transition to homogeneous nuclear matter. In particular, in the limit $\xi\rightarrow0$ all protons would be collected in clusters, while for $\xi=1$ all protons would be free.  The filling factor in our reference model is shown in the left panel of Fig.~\ref{xineffcontour}. As evident from the figure, this does not induce substantial corrections, since $0.9<\xi<1$.
%, as shown in the left panel of Fig.~\ref{xineffcontour}. 

Finally, the proton degeneracy implies a reduction of the number of targets $n_p\rightarrow n_p^{\text{eff}}$ for the Primakoff process. In particular, since protons are fermions the effective number of targets can be calculated by inserting the Pauli blocking factor $(1-f_p)$ in the integral in Eq.~(\ref{np}). Thus $n_p^{\text{eff}}$ results to be~\cite{Payez:2014xsa}
\begin{equation}
n_p^\text{eff}=2\,\xi\int\dfrac{d^3\textbf{p}}{\left(2\pi\right)^3}f_p(1-f_p)\,.
\label{npeff}
\end{equation}
Contours of $n_p^\text{eff}/n_p$ are shown in the right panel of Fig.~\ref{xineffcontour} and indicate a suppression up to $50\%$ in the inner core of the PNS, where the protons are degenerate. 

%%%%%%%%%%%%%%%%%%%%
%%%%%%%%%%%%%%%%%%%%%%%%%%%

%%%%%%%%%%%%%%%
\subsection{Neutrino opacity}
%%%%%%%%%%%%%%%%%%%%%%%%
In order to characterize the axion emissivity in the trapping regime we need to compare the axion opacity with the neutrino one. The {AGILE-BOLTZTRAN} code provides
a detailed characterization of the neutrino opacity (see e.g. \cite{Fischer:2012} for details). However, for the sake of the simplicity, since our treatment of the axion opacity will be simplified we prefer to use also for the neutrino case an approximated recipe. This has the benefit to allows for a semi-analytical calculation of the neutrino opacity.

In the standard scenario
neutrinos are trapped in the SN core and are emitted from the last-scattering surface, the neutrino-sphere at radius $R_{\nu}$.
The strength of the neutrino interactions with matter is characterized through  the opacity $\kappa_\nu$, 
related to the mean-free path $\lambda_\nu$ by $\kappa_\nu\rho=1/\lambda_\nu$. 
Since at early times the neutrino emissivity is dominated by electron species, 
the neutrino opacity can be roughly evaluated  averaging the opacity of electron neutrinos and antineutrinos~\cite{Janka:2000bt}
\begin{equation}
\kappa_{\nu}=\dfrac{L_{\nu_e}\,\kappa_{\nu_e}+L_{\bar{\nu}_e}\,\kappa_{\bar{\nu}_e}}{L_{\nu_e}+L_{\bar{\nu}_e}}\,,
\label{opnu}
\end{equation}
where $L_{\nu_e}$ and $L_{\bar{\nu}_e}$ are the electron neutrino and antineutrino luminosities, while $\kappa_{\nu_e}$ and $\kappa_{\bar{\nu}_e}$ are the electron neutrino and antineutrino opacities, which have contributions from scattering $\kappa_\text{sc}$ and absorption processes $\kappa_\text{ab}$.
%%%%%%%%%%%%%%%%%
%\begin{figure} [t!]
%	\centering
%	\includegraphics[scale=0.45]{ktimes} 
%	\caption{The radial profiles of the neutrino opacity at different post-bounce times $t_{\text{pb}}$. Vertical lines correspond to the neutrino-sphere radius $R_\nu$ at $t_\text{pb}=0.2$~s (continuous black curve), $t_\text{pb}=0.5$~s (red curve) and $t_\text{pb}=1$~s (dashed black curve).}
%	\label{ktimes}
%\end{figure}  
%%%%%%%%%%%%%%%%%%%%%%%%%%%%
 Following the derivation in~\cite{Janka:2000bt}, we define an effective opacity $\kappa_\text{eff}=\sqrt{\kappa_\text{ab}(\kappa_\text{ab}+\kappa_\text{sc})}$ with a schematic expression
\begin{equation}
\begin{aligned}
\kappa_{\nu_e}&\equiv&\kappa_{\text{eff},\nu_e}&&=&&1.62\,\dfrac{\sigma_0\,\left<E_{\nu_e}^2\right>}{m_e^2}\dfrac{1}{m_u}\,X_n\sqrt{1+0.21\dfrac{X_p}{X_n}}\,,\\
\kappa_{\bar{\nu}_e}&\equiv&\kappa_{\text{eff},\bar{\nu}_e}&&=&&1.62\,\dfrac{\sigma_0\,\left<E_{\bar{\nu}_e}^2\right>}{m_e^2}\dfrac{1}{m_u}\,X_p\sqrt{1+0.21\dfrac{X_n}{X_p}}\,,
\end{aligned}
\label{keff}
\end{equation}
where $m_u=1.66\times10^{-24}~\text{g}$ is the atomic mass unit, $m_e=0.511~\text{MeV}$ is the electron rest mass, $\sigma_0=1.76\times10^{-44}~\text{cm}^2$, $X_p$ and $X_n$ are the number fractions of free neutrons and protons. 
%In Fig.~\ref{ktimes} the radial profiles of the neutrino opacity at different times $t_\text{pb}$ are shown. 

%%%%%%%%%%%%%%%
\subsection{The characteristic radii}
%%%%%%%%%%%%%%%%%%%%%%%%
\label{sec:radii}

The features of a SN explosion are strictly connected to three characteristic radii of the SN atmosphere: the neutrino-sphere radius $R_\nu$, the gain radius $R_\text{gain}$ and the shock radius $R_\text{shock}$. 
Note that in multi-dimensional supernova simulations, 
%these quantities represent no 
these quantities do not represent perfect spheres 
%but rather or regions, 
because of the presence of multi-dimensional phenomena such as convection and rotation induced mixing. 
%These definitions are, however, appropriate in a 1 dimensional description of the model. 
The neutrino-sphere radius $R_\nu$ can be evaluated  through the opacity $\kappa_\nu$ in Eq.~(\ref{opnu}) through the relation
%%%%%%%%%%%%%%%%%
\begin{equation}
\tau_\nu(R_\nu)=\int_{R_\nu}^\infty \kappa_{\nu}\,\rho\, dr=\dfrac{2}{3}\,,
\label{defrnu}
\end{equation}
%%%%%%%%%%%%%%%%
where $\tau_\nu$ is the optical depth. This condition corresponds to the requirement that a neutrino emerging from the neutrino-sphere has a probability $e^{-\tau_\nu}=e^{-2/3}$ to reach the infinity. For this reason, in a simplified way the proto-neutron star can be seen as a black-body cooling via neutrino emission from a surface of radius $R_\nu$. 
%The position of  $R_\nu$ at different post-bounce times are shown in  Fig.~\ref{ktimes}.

\begin{figure}[t!!]
	\centering
		\includegraphics[scale=0.33]{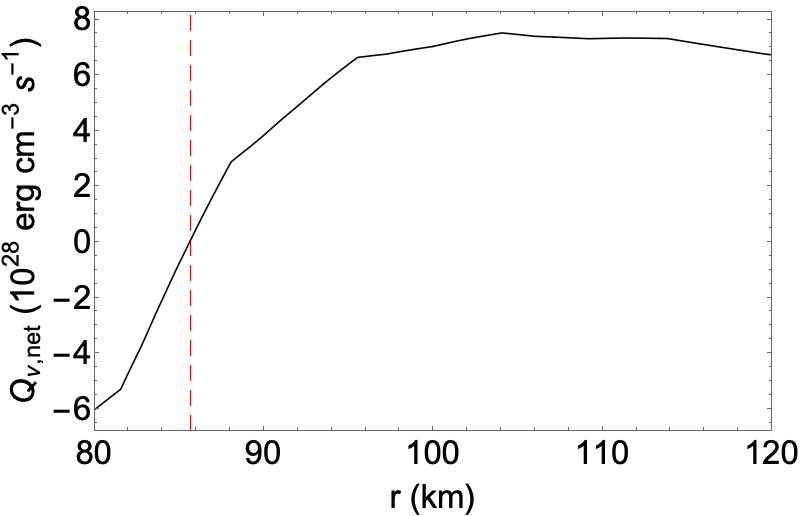} 
	\caption{The net heating rate $Q_{\nu,\text{net}}$ (see text for definition) in the zone where it becomes positive at $t_\text{pb}=0.3$~s. The vertical dashed red line corresponds to $R_\text{gain}\approx86$~km.}
	\label{netrate}
\end{figure}

Actually, neutrino emission and absorption processes determine the cooling and the heating of the matter in the neutrino decoupling region, respectively. In particular, in the SN model of Fig.~\ref{Tempev} charged-current, neutrino-nucleon/nucleus as well as neutrino-electron scattering and neutrino pair processes are considered (see Table~1 in Ref.~\cite{Fischer:2018kdt}, including the updates described in Refs.~\cite{Fischer:2016boc,Fischer:2018kdt}). The total heating rates $Q_\nu$ [in units of MeV~cm$^{-3}$~s$^{-1}$] are defined by the following integral expressions over the neutrino energy, $E$, and relative momentum angle between neutrino propagation and the radial motion, $\mu=\cos\theta$,
\begin{eqnarray}
Q_{\nu_e} &=& \frac{2\pi c}{(h c)^3}
\int dE \,E^3 \int d\mu\,\left\{\kappa_{\nu_e}(E)\,f_{\nu_e}(E,\mu) - j_{\nu_e}(E)(1-f_{\nu_e}(E,\mu))\right\} \\
&+& 
\frac{2\pi}{(h c)^3}
\int dE \,E^2 \int d\mu\, 
\nonumber \\
&& 
\left(
f_{\nu_e}(E,\mu)
\frac{2\pi}{(h c)^3} \int dE' E'^2 \int d\mu' (E-E') R_{\rm scat, \nu_e}^{\rm out}(E,E',\mu,\mu') (1-f_{\nu_e}(E',\mu'))
\right.
\nonumber \\
&&- 
(1-f_{\nu_e}(E,\mu))
\frac{2\pi}{(h c)^3} \int dE' E'^2 \int d\mu' (E-E') R_{\rm scat, \nu_e}^{\rm in}(E,E',\mu,\mu') f_{\nu_e}(E',\mu')
\nonumber \\
&&+
f_{\nu_e}(E,\mu) 
\frac{2\pi}{(h c)^3} \int dE' E'^2 \int d\mu' (E+E') R_{\nu_e\bar\nu_e}^{\rm a}(E,E',\mu,\mu') f_{\bar\nu_e}(E',\mu')
\nonumber \\
&&-
\left.
(1-f_{\nu_e}(E,\mu))
\frac{2\pi}{(h c)^3} \int dE' E'^2 \int d\mu' (E+E') R_{\nu_e\bar\nu_e}^{\rm p}(E,E',\mu,\mu') (1-f_{\bar\nu_e}(E',\mu')
\right)
\nonumber
\nonumber
\end{eqnarray}
of the neutrino emissivity, $j_\nu$ and opacity, $\kappa_\nu$,  in/out-scattering and production/absorption pair processes reaction kernels, $R_{\rm scat, \nu}^{\rm in/out}$ and $R_{\nu\bar\nu}^{\rm p/a}$, taking into account the neutrino phase space occupation in the initial, $f_\nu(E,\mu)$, and final states, $f_\nu(E',\mu')$ (further details can be found Ref.~\cite{Fischer:2012}). The similar expression is obtained for the $\bar\nu_e$ heating rate, $Q_{\bar\nu_e}$, as well as for the heavy-lepton flavors, $Q_{\nu_{\mu/\tau}}$ and $Q_{\bar\nu_{\mu/\tau}}$, except that for the latter two there is no contributions from the charged-current emissivity and opacity. The total net neutrino heating rate is then obtained by summing over all neutrino flavors, denoted as $Q_{\nu, \rm net}=Q_{\nu_e} + Q_{\bar\nu_e} + 2 Q_{\nu_{\mu/\tau}} + 2 Q_{\bar\nu_{\mu/\tau}}$, where the factors of 2 arise since $\mu$- and $\tau$-neutrinos and antineutrinos are treated as one species. Note that contributions with a $+$ sign belong to the total neutrino heating rate, denoted as $Q_{\nu}^+$, while those with a $-$ sign belong to the cooling rate, denoted as $Q_{\nu}^-$. It is useful to define the gain radius, $R_\text{gain}$, as the radius at which the neutrino heating balances the cooling, i.e. the net heating rate $Q_{\nu,\text{net}}\equiv Q_{\nu}^+ - Q_{\nu}^-$ vanishes, as shown in Fig.~\ref{netrate} for $t_\text{pb}=0.3$~s. According to this definition, for $R<R_\text{gain}$ neutrino production prevails, $Q_{\nu}^+ < Q_{\nu}^-$, while at larger distances the absorption processes dominate.
%Therefore one would define $R_\text{gain}$ as the radius at which heating balances cooling, i.e. the net heating rate $Q_{\nu,\text{net}}\equiv Q_{\nu}^+-Q_{\nu}^-$ becomes positive, as shown in Fig.~(\ref{netrate}) for  $t_\text{pb}=0.3$~s.

%%%%%%%%%%%%%%%%%%%%%%%%%%%%%
%%%%%%%%%%%%%%%%%%%%%%%%%%

In the so-called neutrino-driven explosion scenario, the energy deposited at early times ($t_\text{pb}\lesssim 0.3$~s) in the ``gain layer'', i.e. the region between $R_\text{gain}$ and the position of the shock-wave front $R_\text{shock}$, triggers the SN explosion. For this reason it is of crucial importance to evaluate the time evolution of the shock radius $R_\text{shock}$, {beyond which the matter is not yet uncompressed.}
%%%%%%%%%%%%%%%%%%%%%%
%      \begin{figure} [t!!]
%	\centering
%	\includegraphics[scale=0.45]{rshocktimes} 
%	\caption{The radial profiles of the matter density $\rho$ at different post-bounce times $t_{\text{pb}}$. The abrupt drop of the density corresponds to the position of the shock front.}
%	\label{rshocktimes}
%\end{figure}  
%%%%%%%%%%%%%%%%%%%%%%%

\begin{figure} [t!!!]
	\centering
	\includegraphics[scale=0.335]{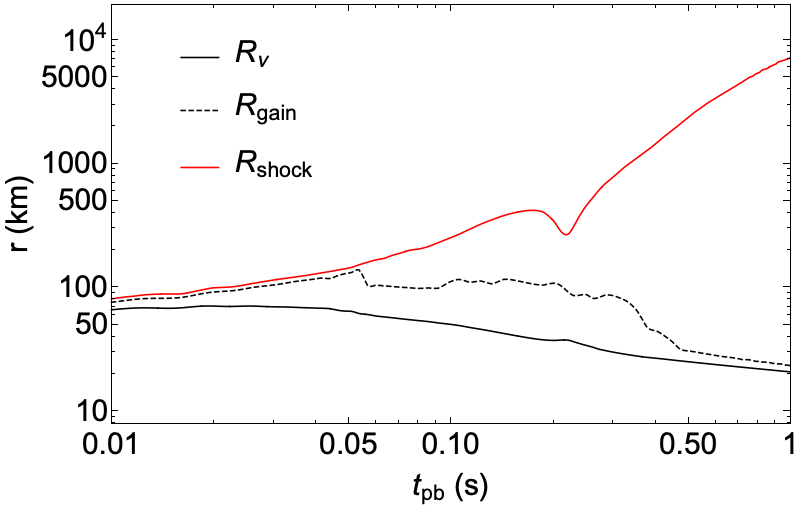} 
	\caption{Time evolution of the neutrino-sphere radius $R_\nu$ (continuous black line), the gain radius $R_\text{gain}$ (dashed black line) and the shock radius $R_\text{shock}$ (continuous red curve).}
	\label{radiitimes}
\end{figure}

In Fig.~\ref{radiitimes} we show the time evolution from $t_\text{pb}=10$~ms to $t_\text{pb}=1$~s of the three characteristic radii. 
We observe that the neutrino-sphere radius (the continuous black curve) is $\sim 50$~km 
at early times ($t_\text{pb}\lesssim100$~ms) and it decreases to $\sim 20$~km at $t_\text{pb}=1$~s. 
The gain radius $R_\text{gain}$ (the dashed black curve) has a similar evolution. 
More specifically, $R_\text{gain}\sim O(100~\text{km})$ in the first 0.3 s after the core-bounce (in agreement with literature estimate~\cite{Janka:2000bt}), but it starts decreasing all the way to $R_\text{gain}\approx23$~km at $t_\text{pb}=1$~s. 
The time evolution of the shock radius $R_\text{s}$ is peculiar of a 1-D simulated SN explosion. 
It increases until $t_\text{pb}\approx200$~ms but at larger times it starts receding. 
After the artificial enhancement of the deposited energy in the gain layer, the SN explosion is triggered 
and thus the shock radius begins to grow again,
%restarts its outward motion, 
reaching $R_\text{shock}\gtrsim O(10^3~\text{km})$ at times larger than $t_\text{pb}\approx0.5$~s.

%%%%%%%%%%%%%%%%%%%%
%%%%%%%%%%%%%%%%%%%%%%%

%%%%%%%%%%%%%%%%%%%%%%%%%%%%%%
\section{ALP emissivity in a supernova}
\label{sec:emiss}
%%%%%%%%%%%%%%%%%%%%%%%%%%%%

%%%%%%%%%%%%%%%%%%%
\subsection{Primakoff process}
%%%%%%%%%%%%%%%%%%

The two-photon coupling of ALPs~\cite{Raffelt:1987im}
%%%%%%%%%%%%
\begin{equation}
\mathcal{L}_{a\gamma\gamma}=-\dfrac{1}{4}g_{a\gamma}a\tilde{F}^{\mu\nu}F_{\mu\nu}\,\ ,
\end{equation}
%%%%%%%%%%%%%%%%%%%%%
allows for ALP production in a SN via the Primakoff process, i.e. the conversion of a photon into an 
ALP in the electric field of nuclei or electrons in the stellar matter.
In the case of massive ALPs the   Primakoff transition rate is given by~\cite{DiLella:2000dn,Cadamuro:2010cz,Carenza:2020zil}
%%%%%%%%%%%%%%%%%%%%
\begin{equation}
\begin{aligned}
\Gamma_{\gamma\rightarrow a} &=g_{a\gamma}^2\dfrac{T\kappa_s^2}{32\pi} \dfrac{p}{E}\bigg\{\dfrac{\left[\left(k+p\right)^2+\kappa_s^2\right]\left[\left(k-p\right)^2+\kappa_s^2\right]}{4kp\kappa_s^2}\ln\left[\dfrac{(k+p)^2+\kappa_s^2}{(k-p)^2+\kappa_s^2}\right]-\\
&-\dfrac{\left(k^2-p^2\right)^2}{4kp\kappa_s^2}\ln\left[\dfrac{(k+p)^2}{(k-p)^2}\right]-1\bigg\}\,,
\end{aligned}
\label{generalrate}
\end{equation}
%%%%%%%%%%%%%%%%%%%%%%%%
where   $p=\sqrt{E^2-m_a^2}$ and $k=\sqrt{\omega^2-{\omega_{\rm pl}}^2}$ are the ALP and photon momentum respectively, 
while $\omega_{\rm pl}$ is the plasma frequency shown in the lower panel of Fig.~\ref{Tempev}, which plays the role of an ``effective photon mass''. We take $E=\omega$ since the energy is conserved. Finally, $\kappa_s$ is an appropriate screening scale which accounts for the finite range of the electric field of the charged particles in the stellar medium. 
In a SN core, the most substantial contribution to the ALP emission via Primakoff comes from free protons. Indeed, electrons are highly degenerate in the SN core. Thus the electron phase space is Pauli-blocked and hence their contribution to the ALP production is negligible. On the other hand, protons are only partially degenerate, as shown in the right panel of Fig.~\ref{meffetacontour}. For this reason, only proton contribution to the Primakoff rate is considered. In the non-degenerate regime, the screening scale would be the Debye one, but in this case, in order to take into account the partial proton degeneracy, the appropriate choice for the inverse screening length is~\cite{Payez:2014xsa}
%%%%%
\begin{equation}
\kappa^2_s=\dfrac{4\,\pi\,\alpha\,n_p^\text{eff}}{T}\,,
\end{equation}
with $n_p^\text{eff}$ given by Eq.~(\ref{npeff}). Note that a larger degeneracy implies the reduction of the effective number density of the targets and therefore the strength of the Primakoff rate is suppressed.

In order to evaluate the energy-loss by Primakoff production, one has to calculate the ALP emissivity $Q_a$, in units of erg cm$^{-3}$ s$^{-1}$, which represents the energy emitted via ALP production per unit volume and time. It results 
%%%%%%%%%%%%
\begin{equation}
Q_a=2\int\dfrac{d^3\textbf{k}}{(2\pi)^3}\Gamma_{\gamma\rightarrow a} \omega f(\omega)=\int_{m_a}^{\infty} dE E \dfrac{d^2n_a}{dtdE}\,,
\label{Qa}
\end{equation}
%%%%%%%%%%%%
where the factor 2 comes from the photon polarization degrees of freedom and $f(\omega)=(e^{\omega/T}-1)^{-1}$ is the Bose-Einstein distribution function of the thermal photons. 
 At fixed mass $m_a$, the emissivity is larger at values of the radius $r$ where the temperature $T$ is higher.
 %%%%%%%%%%%%%%%%%
\begin{figure} [t!!]
	\centering
	\includegraphics[scale=0.34]{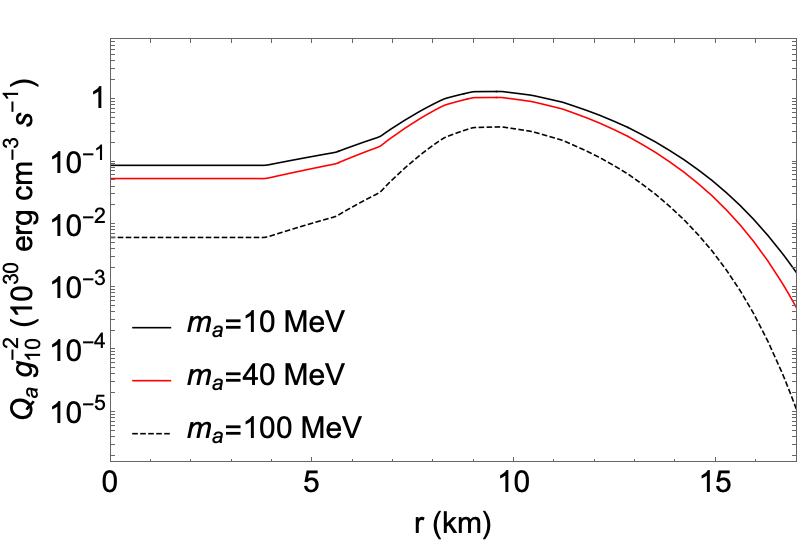}\\
	\caption{ALP emissivity at $t_\text{pb}=1$~s for different values of the ALP mass $m_a$, as shown in legend.}
	\label{Qprim}
\end{figure} 
%%%%%%%%%%%%%%%%%%%%%%%%%%
 In Fig.~\ref{Qprim} we show the ALP emissivity for different values of the ALP mass at $t_\text{pb}=1$~s, normalized to the square of the ALP-photon coupling $g_{10}=g_{a\gamma}/{10^{-10}}$ GeV$^{-1}$. 
It is evident that, regardless of the mass $m_a$, the region of larger production is between $r\sim5-15$~km, 
and the peak of the ALP emissivity is attained at $r\sim10$~km, 
where the temperature reaches its maximum value, as shown in Fig.~\ref{Tempev} (upper left panel). 
It is also noticeable the  Boltzmann suppression of the ALP emissivity, induced by the factor $e^{-m_a/T}$, as the mass increases.
%ALP mass increases, the energy-loss rate decreases because of the.

%%%%%%%%%%%%%%%%%%%%%%%%
\subsection{Photon coalescence}
%%%%%%%%%%%%%%%%%%%%%%%%

In a medium of sufficiently high density, 
ALPs can also be produced through the so-called ``photon coalescence'' 
or ``inverse decay process''~\cite{DiLella:2000dn}, where two photons can annihilate producing an axion.
%two photons can annihilate producing an axion, in the so called ``photon coalescence''  or ``inverse decay process''~\cite{DiLella:2000dn}. 
This process has a kinematic threshold, vanishing for $m_a<2 \omega_{\rm pl}$. 

\begin{figure} [t!]
	\centering
	{\includegraphics[scale=0.34]{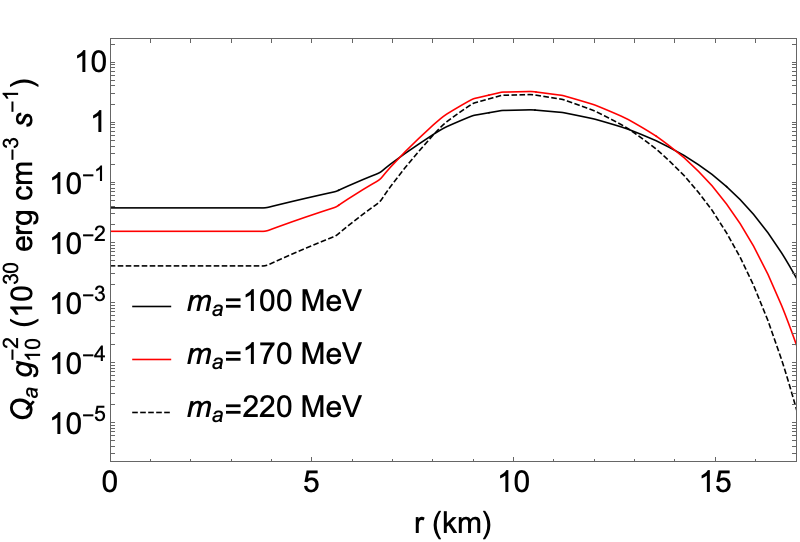} 
	}  
	\caption{ ALP emissivity from photon coalescence  for different values of the ALP mass at $t_\text{pb}=1$~s. 
	}
	\label{qinv}
\end{figure}

In order to evaluate the axion production rate from photon coalescence in a thermal medium, 
it is convenient to approximate the Bose-Einstein photon distribution 
with a Maxwell-Boltzmann $f(E)\rightarrow e^{-E/T}$ for the photon occupation number~\cite{DiLella:2000dn}. 
The approximation is well justified since, at masses for which the coalescence process dominates, $m_a\gtrsim 100~{\rm MeV}$, $E\geq m_a\gg T$.
Thus, the production rate per unit volume and energy can be expressed as~\cite{Carenza:2020zil}
\begin{equation}
\dfrac{d^2n_a}{dtdE}=g_{a\gamma}^2 \dfrac{m_a^4}{128\pi^3}p\left(1-\dfrac{4\omega_{\rm pl}^2}{m_a^2}\right)^{3/2}e^{-E/T}\,,
\label{dndeinverse}
\end{equation}
with $p=\sqrt{E^2-m_a^2}$. 
The axion emissivity can be calculated as before 
\begin{equation}
Q_a=\int_{m_a}^{\infty} dE E \dfrac{d^2n_a}{dtdE}\,.
\end{equation}
At fixed value of the axion mass $m_a$, the emissivity is larger at radii $r$ where the temperature is higher because of the Boltzmann factor in Eq.~(\ref{dndeinverse}). 
The ALP emissivities for different values of the mass $m_a$ at $t_\text{pb}=1$~s are shown in Fig.~\ref{qinv}. 
Regardless of the axion mass, the emissivity is maximal at $r\approx10$~km and the production region 
via  photon coalescence is the same as Primakoff, between $r\approx5-15$~km. 
As depicted in Fig.~\ref{qinv}, in the production region the emissivity increases 
until $m_a\approx170$~MeV and then it starts decreasing because of the Boltzmann suppression, 
while outside this region the ALP production is strongly suppressed for any mass value. 
%%%%%%%%%
%%%%%%%%%%%%%%

\begin{figure} [t!!]
	\centering
	\includegraphics[scale=0.35]{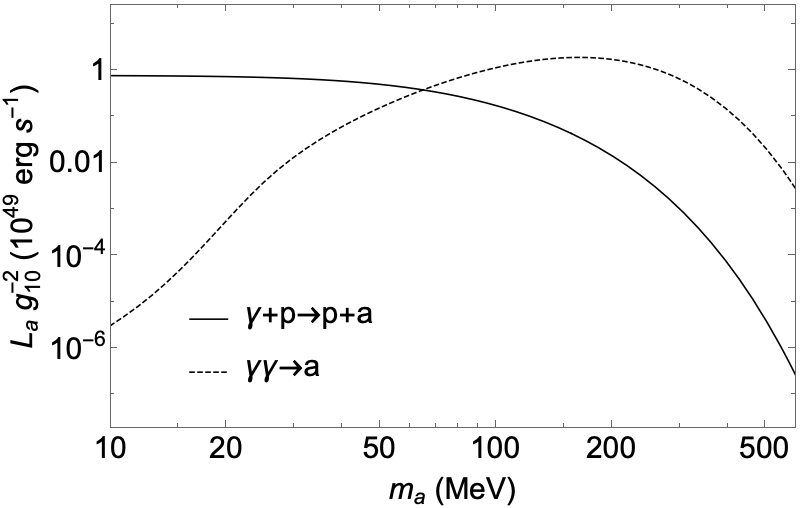}   \\
	\caption{ALP luminosity for Primakoff (continuous curve) and photon coalescence (dashed curve) as a function of the axion mass $m_a$ at $t_\text{pb}=1$~s.}
	\label{lumcfr}
\end{figure}

Finally, the ALP luminosity, i.e. the energy emitted per unit time (measured in erg s$^{-1}$), is given integrating  the emissivity over the SN model, i.e. 
%%%%%%%%%%%%%%%%%%%%%%%%%%%
\begin{equation}
L_a=4\pi\int Q_a(r) r^2 dr\,.
\label{luma}
\end{equation}
%%%%%%%%%%%%%%%%%%%%%%%%
In Fig.~\ref{lumcfr} the axion luminosities from Primakoff and photon coalescence as a function of the axion mass $m_a$ at $t_\text{pb}=1$~s are represented. It is apparent that the coalescence process is sub-leading for $m_a\lesssim70$~MeV, while at larger masses it becomes dominant and reaches its maximum at $m_a\approx170$~MeV. 
%%%%%%%%%%%%%%%%%%%%%%
%%%%%%%%%%%%%%%%%%%%%

%%%%%%%%%%%%%%%%%%%%%%%%%
\section{SN 1987A ALP bounds}
\label{sec:bound}
%%%%%%%%%%%%%%%%%%%%%%%%%%

%%%%%%%%%%%%%%%%%%%%%%%%%%%
\subsection{Modified luminosity criterion}
%%%%%%%%%%%%%%%%%%%%%
The SN 1987A neutrino observations by
 KII and IMB experiments are in good
agreement with the standard picture of the proto-neutron star cooling by neutrinos 
on a time scale of ${\mathcal O}(10~\rm{s})$ (see, e.g.,~\cite{Pagliaroli:2008ur}).
If ALPs are able to transport energy out from the interior of the PNS, they would provide a new efficient cooling mechanism. Observationally, this implies that the cooling time scale would be shortened. In particular, the observed duration of the neutrino signal implies that the luminosity carried away by ALPs from the interior of the PNS to the outside of the neutrino-sphere, $L_a$, must not exceed the neutrino luminosity in all the six (anti)neutrino degrees of freedom $L_\nu$ in the cooling phase. This is the so-called ``energy-loss argument''. Conventionally, it is taken as benchmark the neutrino luminosity value at $t_\text{pb}=1$~s,
  $L_\nu(t_\text{pb} \sim 1~\text{s})\simeq 3\times 10^{52}\,\text{erg s}^{-1}\,$. Thus, the constraint on ALP emissivity is 
  obtained requiring~\cite{Raffelt:2006cw}
  %%%%%%%%%%%%%%%%%%
  \begin{equation}
L_a(t_\text{pb}=1~\text{s})\lesssim 3\times 10^{52}\,\text{erg s}^{-1}\,.
\label{bound}
\end{equation}
We stress that several numerical simulations have shown that the duration of the neutrino burst would be roughly halved when the limit~(\ref{bound}) is saturated~\cite{Raffelt:2006cw}.
  \newline
As recently proposed in~\cite{Chang:2016ntp,Ertas:2020xcc}, only the ALP luminosity that cannot be reprocessed efficiently as neutrino energy is relevant to constrain the ALP parameter space. 
More precisely, if ALPs are produced in the zone of neutrino diffusion, i.e. behind the neutrino-sphere, energy is taken away from there. 
However, if they are absorbed in a region where the neutrino production is still efficient, 
the energy they deposit could be re-emitted via neutrino production and the neutrino signal would result to be practically unaltered. 
On the other hand, if ALPs reach a radius $R_\text{far}>R_\nu$ beyond which the neutrino production is negligible, 
the deposited energy would result to be essentially unavailable to them. 
For this reason, 
%closely following the derivation given in~\cite{Chang:2016ntp}, 
%\mg{I think we can remove the sentence "closely following the derivation given in~\cite{Chang:2016ntp}" since it is clear we are following that reference here}
the axion luminosity can be evaluated by including an optical depth factor characterizing the probability 
that an ALP produced in the core region ($r\lesssim R_\nu$) reaches the radius $R_\text{far}$. 
There is a number of reasonable choices for $R_\text{far}$, 
the only stringent condition is that $R_\text{far}>R_\nu$. 
A lower bound on $R_\text{far}$ is the neutrino gain radius $R_\text{gain}$, 
outside of which the neutrino production has a lower rate than the absorption one. 
On the other hand, a reasonable upper limit is the shock radius $R_\text{shock}$, outside of which matter is not yet uncompressed. 
In this work, we fix $R_\text{far}=R_\text{gain}$. 
From our simulation at $t_\text{pb}=1$~s we find $R_\text{gain}\approx 23$~km, as shown in Fig. \ref{radiitimes}.

For a fixed value of $R_\text{far}$, the ALP luminosity $L_a$ results to be~\cite{Chang:2016ntp}
%%%%%%%%%%%
\begin{equation}
L_a=4\pi \int_0^{R_\nu} dr r^2 \int_{m_a}^{\infty} dE E \dfrac{d^2 n_a}{dtdE} e^{-\tau_a(r,E,R_\text{far})}\,,
\label{lumnuova}
\end{equation}
%%%%%%%%%%%%%%
where the volume integration is till $R_\nu$ since we are interested just in the energy taken away from behind the neutrino-sphere and the ALP volume emission rate per unit energy  is given by Eq.~(\ref{gravtrapcond}), while $e^{-\tau_a(r,E,R_\text{far})}$ is the optical depth factor which takes into account the absorption effects. In particular, $\tau_a(r,E,R_\text{far})$,
the optical depth of an ALP produced at $r$ with energy $E$ reaching $R_\text{far}$, results to be~\cite{Chang:2016ntp}
%%%%%%%%%%%%%%%%%%
 \footnote[2]{The pre-factor $\left(1-\dfrac{r(r-R_\text{c})}{2R_\nu^2}\right)$ takes into account the non-radial trajectories, but it does not deviate substantially from one \cite{Chang:2016ntp}.} 
%%%%%%%%%%%%%%
\begin{equation}
\tau_a(r,E,R_\text{far})=\left(1-\dfrac{r(r-R_\text{c})}{2R_\nu^2}\right)\int_r^{R_\text{far}}  \dfrac{d\tilde{r}}{\lambda_a(E,\tilde{r})}\,,
\end{equation}
%%%%%%%%%%%%%%%%%%%%%%%
where $R_\text{c}\approx10$~km is the core radius and $\lambda_a$ is the total axion mean free path (mfp)
%%%%%%%%%%%%%%%
\begin{equation}
\lambda_a^{-1}=\lambda_{a\rightarrow\gamma}^{-1}+\lambda_{a\rightarrow\gamma\gamma}^{-1}\,,
\end{equation}
with $\lambda_{a\rightarrow\gamma}$ the inverse Primakoff mfp and $\lambda_{a\rightarrow\gamma\gamma}$ the decay mfp. In particular, the decay mfp $\lambda_{a\rightarrow\gamma\gamma}$ is
%%%%%%%%%%%%%%%%%%%
\begin{equation}
\lambda_{a\rightarrow\gamma\gamma}=\dfrac{\beta_E\,\gamma_E}{\Gamma_{a\rightarrow\gamma\gamma}}\,,
\label{mfpdec}
\end{equation}
%%%%%%%%%%%%%%%%%%%
where $\gamma_E=E/m_a$ is the Lorentz factor, $\beta_E=\sqrt{1-\gamma_E^{-2}}$ and $\Gamma_{a\rightarrow\gamma\gamma}$ is the decay rate 
%%%%%%%%%%%%%%%%%
\begin{equation}
\Gamma_{a\rightarrow\gamma\gamma}=g_{a\gamma}^2\dfrac{m_a^3}{64\pi}\left(1-\dfrac{4\omega_{\rm pl}^2}{m_a^2}\right)^{3/2}\,.
\end{equation}
%%%%%%%%%%%%%%%
On the other hand, the inverse Primakoff mfp $\lambda_{a\rightarrow\gamma}$ results to be
%%%%%%%%%%%%
\begin{equation}
\lambda_{ a\rightarrow\gamma}=\dfrac{\beta_E}{\Gamma_{a\rightarrow\gamma}}\,,
\label{mfpprim}
\end{equation}
%%%%%%%%%%%%%%%%
where
$\Gamma_{a\rightarrow\gamma}=({2\beta_\gamma}/{\beta_E})\Gamma_{\gamma\rightarrow a}$ 
is the inverse Primakoff rate,
 %$\Gamma_{a\rightarrow\gamma}=({2\beta_\gamma}/{\beta_E})\Gamma_{\gamma\rightarrow a}$, 
 with $\Gamma_{\gamma\rightarrow a}$ given by Eq.~(\ref{generalrate}),
 and the factor $2$ accounting for the photon polarization. 

Since the integration over the energy in Eq.~(\ref{lumnuova}) influences the optical depth $\tau_a$, for computational reasons we decided to define a mean optical depth $\tau_a^*$
%%%%%%%
\begin{equation}
\tau_a^*(r,R_\text{far})\equiv\tau_a(r,\langle E_a \rangle,R_\text{far})
\end{equation}
%%%%%%%%
where $\langle E_a \rangle$ is the average axion energy over the emission spectrum. \newline
Therefore the luminosity can be rewritten as
%%%%%%%%%%%%%
\begin{equation}
L_a=4\pi\int_0^{R_\nu} dr\,r^2 e^{-\tau_a^*(r,R_\text{far})}\int_{m_a}^{\infty} dE E \dfrac{d^2 n_a}{dtdE}\equiv4\pi\int_0^{R_\nu} dr\,r^2 Q_a(r)e^{-\tau_a^*(r,R_\text{far})}\,,
\label{lumax}
\end{equation}
%%%%%%%%%%%%%%%%%%
which is exactly the same expression in Eq.~(\ref{luma}) modified with the inclusion of the optical depth factor $e^{-\tau_a^*}$. Thus the luminosity is essentially determined by the product of two factors: the emissivity $Q_a$, taking into account the axion production processes, and the optical depth factor $e^{-\tau_a^*}$, representing the axion absorption processes. Since both $Q_a$ and $\tau_a$ increases as $g_{a\gamma}^2$, the luminosity $L_a$ depends on the coupling constant $g_{a\gamma}$ as $L_a\sim g_{a\gamma}^2\,e^{-g_{a\gamma}^2}$. For this reason at fixed value of the axion mass $m_a$, there are two critical values of the coupling $g_{a\gamma}$: $g_{a\gamma}^{\text{L}}$ and $g_{a\gamma}^{\text{H}}$, where the superscripts ``L'' and ``H'' stand respectively for ``low'' and ``high''. 
For $g_{a\gamma}<g_{a\gamma}^{\text{L}}$, ALPs are so weakly coupled that they cannot be produced readily enough to affect the evolution of the PNS. On the other hand, for $g_{a\gamma}>g_{a\gamma}^{\text{H}}$ ALPs are trapped before they reach $R_\text{far}$, allowing the deposited energy to be efficiently reconverted in the form of thermal neutrinos.
%%%%%%%%%%%%
\begin{figure}[t!!]
	\centering
	\includegraphics[scale=0.45]{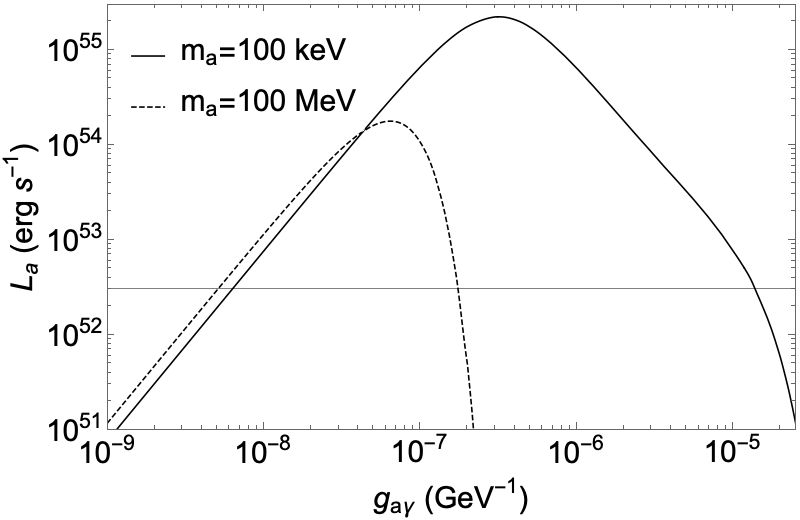}   \\
	\caption{ALP luminosity in Eq.~(\ref{lumax}) as a function of the coupling constant $g_{a\gamma}$ for $m_a=100$~keV (continuous black curve) and $m_a=100$~MeV (dashed black curve), evaluated at $t_\text{pb}=1$~s with $R_\text{far}=R_\text{gain}\approx23$~km. The horizontal continuous black line is the critical value $L_\nu=3\times10^{52}$~erg~s$^{-1}$.}
	\label{lumaxintersect}
\end{figure} 
%%%%%%%%%%%%%
 As shown in Fig.~\ref{lumaxintersect}, for all the values in the range $g_{a\gamma}^{\text{L}} \leq g_{a\gamma}\leq g_{a\gamma}^{\text{H}}$ the axion luminosity $L_a$ violates the bound $L_a \lesssim L_\nu$ and thus these values must be excluded. 

\subsection{Free-streaming regime}
\begin{figure} [t!!]
	\centering
	\includegraphics[scale=0.44]{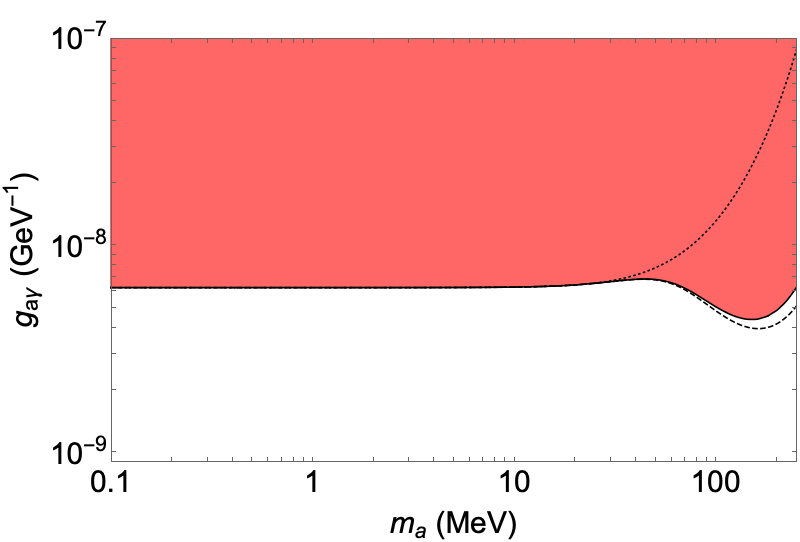}   
	\caption{ALP exclusion plot in the $g_{a\gamma}-m_a$ plane coming from the energy-loss argument. The dotted black curve represents the bound accounting
	the ALP emissivity by only Primakoff process while the dashed black curve includes also the photon coalescence process. The continuous black curve includes the gravitational trapping effect, which is ignored in the other curves. The red region is excluded by the energy-loss argument in the free-streaming regime by accounting for the gravitational trapping.}
	\label{freestreamplot}
\end{figure}

In the small coupling limit [$g_{a\gamma} \lesssim O(10^{-8}~\rm{GeV}^{-1})$], the optical depth $\tau_a^*(r,R_\text{far})\ll1$ and thus $e^{-\tau_a^*(r,R_\text{far})}\sim1$. Physically, this means that the axion mfp is significantly larger than the scale of the PNS, therefore ALPs can free stream as soon as they are produced in the SN core, leading to volume emission. In this context, the expression for the axion luminosity $L_a$ reduces to Eq.~(\ref{luma}). Imposing that $L_a$ satisfies Eq.~(\ref{bound}), one gets a constraint on the ALP-photon coupling $g_{a\gamma}$ as a function of the ALP mass $m_a$. Our result is shown in the exclusion plot
reported in Fig.~\ref{freestreamplot}, in which the dotted line represents the bound obtained assuming only the Primakoff process, while the dashed curve takes into account both the Primakoff and the photon coalescence contributions. In the low-mass limit ($m_a\lesssim$ a few MeV), where the photon coalescence is not relevant, values of the ALP-photon coupling $g_{a\gamma}\gtrsim 6\times 10^{-9}$~GeV$^{-1}$ are excluded, in agreement with previous results~\cite{Lee:2018lcj}. 
A comparison between the dotted curve and the dashed one reveals that for masses above a few 10~MeV the photon coalescence is no longer negligible. 
Indeed, including this channel the bound is strengthened by a factor $\gtrsim 3$ for $m_a\gtrsim100$~MeV 
and by over an order of magnitude for $m_a\gtrsim 200$~MeV. 
We note that for masses $m_a \gtrsim 170$~MeV the bound is weakened since the ALP production is Boltzmann-suppressed.

\subsection{Gravitational trapping}
\label{subsec:gravtrap}
ALPs produced with a kinetic energy satisfying 
\begin{equation}
E_\text{kin}\leq K_\text{tr}\equiv \dfrac{G_\text{N} M_r m_a}{r}\,,
\label{gravtrap}
\end{equation}
will not free stream, since they are trapped by gravitational attraction~\cite{Dreiner:2003wh}. In Eq.~(\ref{gravtrap}), $G_\text{N}$ is the Newton constant, $r$ is the radius at which the ALP is produced and $M_\text{r}$ is the mass of supernova enclosed within the radius $r$. One can schematically include this effect by modifying the ALP volume emission rate per unit energy as
\begin{equation}
\dfrac{d^2n_a}{dEdt}\rightarrow \dfrac{d^2n_a}{dEdt}\,\theta(E-m_a-K_\text{tr})\,.
\end{equation}
At $t_\text{pb}=1$~s, the gravitational potential $U_G(r)=\dfrac{G_\text{N} M_r}{r}$ has a maximum 
at $r_\text{max}\simeq17$~km, where $U_G(r_\text{max})=U_\text{max}\simeq0.12$. 
Therefore ALPs produced at $r\leq r_\text{max}$ must have a kinetic energy
$E_\text{kin}>m_a U_\text{max}$ in order to escape from the potential well, otherwise they will be gravitationally trapped. For this reason the effect of the gravitational trapping is accounted in the ALP emissivity as
%%%%%%%%%%%%%%%%%%%%%
%.................................................................
\begin{equation}
\label{gravtrapcond} \dfrac{d^2n_{a,tr}}{dEdt}  = \left\{
\begin{array}{ll}
 \dfrac{d^2n_a}{dEdt}\,\theta(E-m_a-m_a U_\text{max}) & \text{if}~r\leq r_\text{max} \,\ , \\
  & \\
\dfrac{d^2n_a}{dEdt}\,\theta(E-m_a-K_\text{tr})   &\text{if}~r> r_\text{max} \,\ . \\
 \end{array}\right.
\end{equation}
%.....................................................................
With this prescription we can compute the emissivity $Q_a$ in Eq.~(\ref{Qa}) and the luminosity $L_a$ in Eq.~(\ref{lumax}).
%%%%%%%%%%%%%%%
\begin{figure}[t!!]
	\centering
	\includegraphics[scale=0.44]{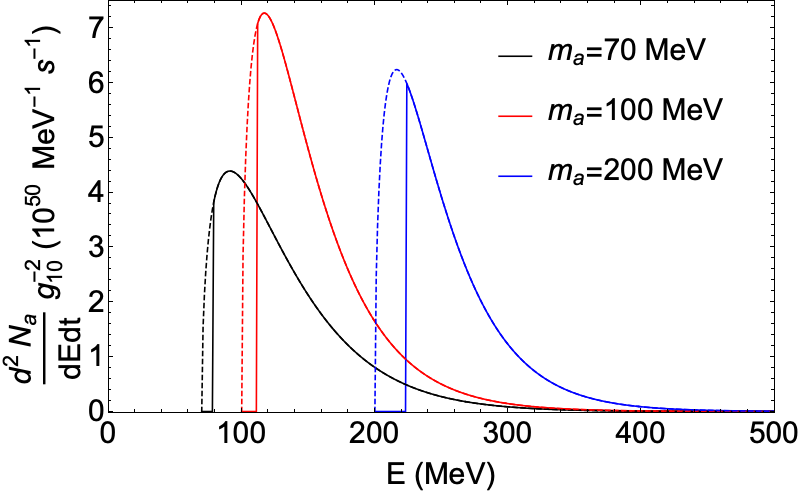}   
	\caption{Total ALP production rate per unit energy at $t_\text{pb}=1$~s for $m_a=70$~MeV (black curves), $m_a=100$~MeV (red curves) and $m_a=200$~MeV (blue curves). The dashed curves refer to cases without the inclusion of  the gravitational trapping effect, which is accounted for in the continuous curves.}
	\label{dNgravtrap}
\end{figure} 
%%%%%%%%%%%
 We stress that, since at $t_\text{pb}=1$~s the ALP production region is in the region  $r\in [5;15]$~km, essentially all the ALPs must satisfy the condition in the first line of Eq.~(\ref{gravtrapcond}). This implies that at fixed mass $m_a$, the total ALP emission rate per unit energy would be cut for $E_a\lesssim 1.12\,m_a$ because of the gravitational attraction, as shown in Fig.~\ref{dNgravtrap}. It is evident that this effect becomes more important as the ALP mass $m_a$ increases, thus the energy-loss bound results to be relaxed at larger masses, as depicted in Fig.~\ref{freestreamplot}. In particular, for $m_a\lesssim100$~MeV the gravitational trapping is negligible, while the bound is relaxed by $\gtrsim 15\%$ for $m_a\gtrsim200$~MeV.%, as shown in Fig.~\ref{figgravtrap} (lower panel).

%%%%%%%%%%%%%%
%\begin{figure}[t!]
%	\centering
%	\includegraphics[scale=0.41]{boundgravtrap}   \\
%	\caption{ALP exclusion plot  from the energy-loss argument in the plane $g_{a\gamma}-m_a$. The continuous black curve includes 
%	 the gravitational trapping effect, which is ignored in the dash-dotted black curve.}
%	\label{boundgravtrap}
%\end{figure} 
%%%%%%%%%%%%%%%%%%%
%%%%%%%%%%%%%%
%\begin{figure}[t!]
%	\centering
%	\includegraphics[scale=0.41]{discrepancy}   \\
%	\caption{The discrepancy between the bound on the coupling $g_{a\gamma}$ evaluated accounting the gravitational trapping effect and the one obtained by ignoring this effect.}
%	\label{discrepancy}
%\end{figure} 
%%%%%%%%%%%%%%%%%%%

%\begin{figure}[t!]
%	\centering
%		\includegraphics[scale=0.42]{boundgravtrap}   \\
%		\vspace{20 pt}
%	\includegraphics[scale=0.42]{discrepancy}   \\
%	\caption{
%		\textit{Upper panel}: ALP exclusion plot  from the energy-loss argument in the plane $g_{a\gamma}-m_a$. The continuous black curve includes 
%		the gravitational trapping effect, which is ignored in the dash-dotted black curve. \textit{Lower panel}: The discrepancy between the bound on the coupling $g_{a\gamma}$ evaluated accounting the gravitational trapping effect and the one obtained by ignoring this effect.}
%	\label{figgravtrap}
%\end{figure} 
%%%%%%%%%%%%%%%%%%%

%%%%%%%%%%%%%%%%%%%%%
\subsection{Trapping regime}
\begin{figure}[t!]
	\centering
	\includegraphics[scale=0.44]{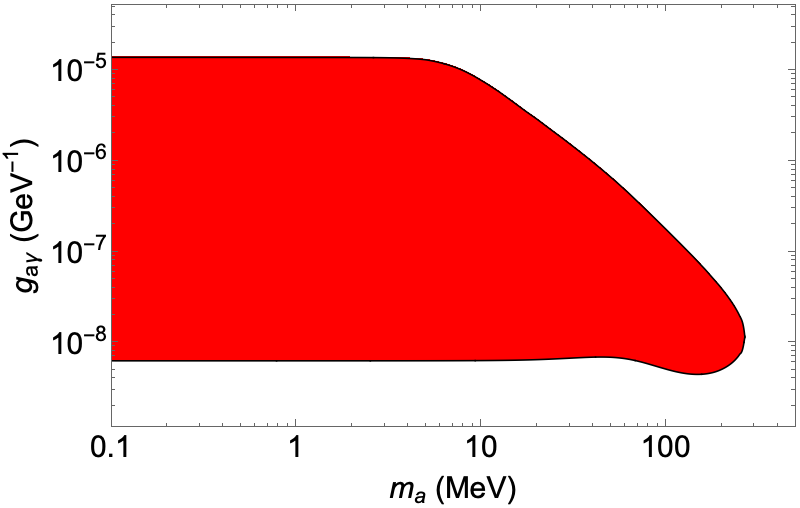}   \\
	\caption{ALP exclusion plot in the $g_{a\gamma}-m_a$ plane obtained through the modified luminosity criterion.}
	\label{boundmodlum}
\end{figure}

%%%%%%%%%%%%%%%%%%%%
As the ALP-photon coupling $g_{a\gamma}$ increases, ALPs produced in the SN core interact so strongly that their  mean free path becomes smaller than the size of the SN core ($R_{\rm{c}}\sim 10$~km). In this context, the most ALPs produced within the neutrino-sphere do not actually escape but they are reabsorbed: this is the so-called ``trapping regime''. In this regime, the most relevant factor in Eq.~(\ref{lumnuova}) is the optical depth factor, characterizing the probability that an ALP produced within the neutrino-sphere reaches $R_{\rm{gain}}$. The red region shown in Fig.~\ref{boundmodlum} is excluded through the modified luminosity criterion Eq.~(\ref{bound}). For each value of the axion mass, the lower bound $g_{a\gamma}^L$ is obtained in the free-streaming regime, while the upper bound $g_{a\gamma}^H$ in the trapping regime. It is interesting to stress that at $m_{a}\approx266$ MeV the lower bound and the upper one are smoothly connected since the transition from the free-streaming to the trapping regime is naturally described by the optical depth factor itself. In the trapping regime, in the small mass limit ($m_{a}\lesssim10$ MeV), values of the coupling $g_{a\gamma}\lesssim 1.38\times10^{-5}$ GeV$^{-1}$ are excluded, while for larger masses the bound is relaxed.\newline
%%%%%%%%%%%%%%
%%%%%%%%%%%%%%%%%%%
A different strategy, based on the axion opacity, was often used in literature to constrain the ALP parameter space in the trapping regime. In this context, in the small mass limit ($m_a\lesssim 10$~MeV), where the dominant absorption process is the inverse Primakoff process, axions are supposed to be thermally emitted from the so-called axion-sphere, the analogous of the neutrino-sphere. This assumption has been critically assessed in Ref.~\cite{Chang:2016ntp}, which shows that assuming a thermal spectrum underestimates the luminosity and thus leads to weaker limits. On the other hand, at larger masses the axion-sphere would not be well defined since the dominant absorption process is the ALP decay (see Appendix~\ref{app:trapmethods} for details). For these values of the mass, by imposing that the energy transferred by ALPs in the SN core must be smaller than the one transported by neutrinos, one would obtain a bound on the ALP-photon coupling stronger than the one obtained through the modified luminosity criterion. A detailed comparison of these different criteria is performed in Appendix \ref{app:trapmethods}. \newline
Finally, we explored the dependence on the SN model of the bound on $g_{a\gamma}$. As shown in Appendix \ref{app:uncertainties}, where we re-evaluate the bound using a SN model with a progenitor mass of 25.0 $M_\odot$, a larger progenitor mass implies a larger exclusion region in the ALP parameter space due to the higher temperature in the SN core. However, independently on the progenitor mass, no SN simulation takes into account the axion feedback to the SN explosion, thus a self-consistent inclusion of ALPs in a SN simulation is necessary to have a more reliable bound on the ALP parameter space. Performing such a simulation would be a challenging task (see, e.g., Ref.~\cite{DeRocco:2019fjq} for a recent investigation in the context of dark photons), and demands a separated investigation.

%%%%%%%%%%%%%%%%%%%%%%
\section{Shock revival and ALP energy deposition}
\label{sec:endep}
%%%%%%%%%%%%%%%%%%%%%

Due to the photo-dissociation of heavy nuclei the SN shock wave loses its strength and after  $t_{\rm pb }\sim100$~ms it stalls and would eventually fall back on the SN core, if it is not revitalized by some energy injection. In the ``neutrino-driven explosion scenario,'' the shock is revived by neutrino heating aided by multidimensional hydrodynamical effects~\cite{Muller:2017hht}, ultimately leading to a SN explosion. However, in one-dimensional simulations the heating rates for neutrino reactions are artificially  increased inside the heating region to trigger the explosion. Here we investigate the intriguing possibility that the ALP production in the SN core and their subsequent decay inside the mantle would heat the SN matter and increase the total energy of the envelope, helping the revival of the shock and the trigger of the explosion even in one dimensional simulations. Indeed, ALPs decaying into photons would provide a pressure gradient and an energy deposition in the region behind the shock, since photons quickly thermalize with matter. An amount of energy $E_\text{dep}$ deposited in a region with mass $M$ and temperature $T$ would give an increase in entropy-per-mass~\cite{Fuller:2009zz}
%%%%%%%%%
\begin{equation}
\Delta s \approx \dfrac{E_\text{dep}}{TM/m_u}\,,
\label{deltas}
\end{equation} 
%%%%%%%%%%%
where $m_u$ is the atomic mass unit.  As the entropy-per-baryon increases, nuclei (partially) melt and at least some of the photo-dissociation burden on the shock would be relieved, definitely helping the trigger of the explosion.

 A comparison between the neutrino heating in the gain layer and the ALP one would be an interesting starting point to assess the impact of ALPs on the explosion. In particular, if the energy deposited by ALPs in the gain layer competes with the neutrino one before the explosion is artificially triggered ($t_\text{pb}\lesssim 250$~ms), the contribution of the decaying ALPs would help the revival of the shock. 
 
At each time step, we evaluate the rate of energy deposited by neutrinos in the gain layer as
%%%%%%%%
\begin{equation}
L_{\nu,\,\text{gain}}(t)=4\pi\int_{R_\text{gain}}^{R_\text{shock}} dr\,r^2\,(Q^+_\nu-Q^-_\nu)\,,
\end{equation}
%%%%%%%%%%%%%%%%%%
where 
$Q^+_\nu$ and $Q^-_\nu$ are respectively the heating and the cooling rate per unit volume, $R_\text{gain}$ is the gain radius and $R_\text{shock}$ the shock radius, described in Sec.~\ref{sec:radii}. By integrating the rate $L_{\nu,\,\text{gain}}$ over time, the energy deposition until the time $t$ is obtained
\begin{equation}
E_{\nu,\,\text{dep}}(t)=\int_{t_0}^{t} d\tilde{t}\, L_{\nu,\,\text{gain}} (\tilde{t})\,,
\end{equation}
where we fix $t_0=10$~ms since we are interested in the shock propagation after the neutronization burst.
%%%%%%%%%%%%%%%%%%
    \begin{figure}[t!]
	\centering
		\includegraphics[scale=0.4]{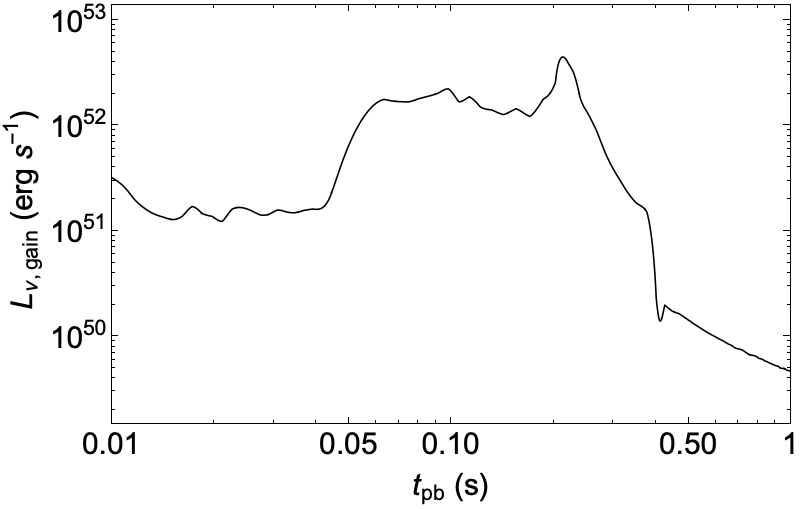}
		\\
		\vspace{8 pt}
		\includegraphics[scale=0.4]{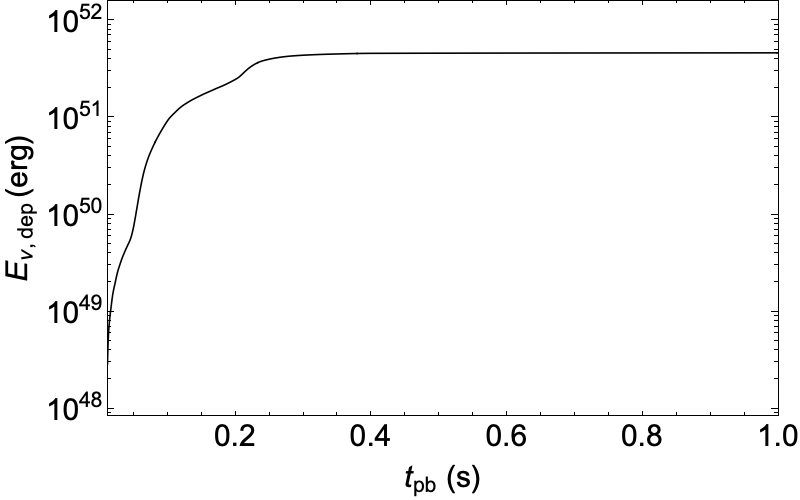} 
	\caption{\textit{Upper panel}: The neutrino energy deposition rate in the gain layer in the time-window $t_\text{pb}\in[10^{-2},1]$~s. \textit{Lower Panel}: The energy deposited by neutrinos in the gain layer in the time-window $t_\text{pb}\in[10^{-2},1]$~s.}
	\label{ennu}
\end{figure} 
%%%%%%%%%%%%%%%%
As shown in the upper panel of Fig.~\ref{ennu}, the neutrino energy deposition rate is $L_{\nu,~\text{gain}}\sim O(10^{51}-10^{52}~\text{erg~s}^{-1})$ for times $t_{\text{pb}}\lesssim 0.2$~s. The bump at $t_\text{pb}\approx220$~ms corresponds to the artificial enhancement of the neutrino luminosity to trigger the explosion, while after the explosion has set up, $L_{\nu,\,\text{gain}}$ starts to decrease, becoming smaller than $O(10^{50}~\text{erg s}^{-1})$ at $t_\text{pb}=1$~s. This implies that the energy deposited by neutrinos in the gain layer increases until $t_\text{pb}\approx0.3$~s, saturating at $E_\text{dep}\approx 5\times 10^{51}$~erg s$^{-1}$ for larger times, as shown in the lower panel of Fig.~\ref{ennu}. \newline
In order to evaluate the ALP heating, we focus our attention on the mass range $100$ MeV $\leq m_a \leq 300$ MeV, where the dominant processes are the production via photon coalescence and the absorption via decay. At each time step, we evaluate the axion luminosity through Eq.~(\ref{luma}). Assuming for the sake of simplicity that ALPs are produced at a mean radius $R_{\text{p}}=\left(\int dr \,r\, Q_a(r)\right)/\int dr\, Q_a(r)$,
 the rate of energy deposited at a distance $R$ results to be
\begin{equation}
L_{a,\,\text{dep}}(t,R)=L_a(t)\left[1-\exp\left(-\int_{R_p}^{R}  \dfrac{dr}{\lambda_{a\rightarrow\gamma\gamma}(\left<E_a\right>,r)}\right)\right]\,,
\end{equation}
with $\lambda_{a\rightarrow\gamma\gamma}$ given by Eq.~(\ref{mfpdec}) and $\left<E_a\right>$ the average axion energy over the emission spectrum. 
The rate at which the ALP energy is deposited in the gain layer results to be
%%%%%%%%%%%%%%%
\begin{equation}
L_{a,\,\text{gain}}(t)=L_{a,\,\text{dep}}(t,\,R_\text{shock})-L_{a,\,\text{dep}}(t,\,R_\text{gain})\,,
\end{equation}
and the deposited energy is obtained by integrating over time
\begin{equation}
E_{a,\,\text{dep}}=\int_{t_0}^{t} d\tilde{t}\, L_{a,\,\text{gain}} \left(\tilde{t}\right)\,.
\end{equation}
%%%%%%%%%%%%%% 
 %%%%%%%%%%%%%%%%%%%%% 
     \begin{figure}[h!]
 	\centering
 		\includegraphics[scale=0.185]{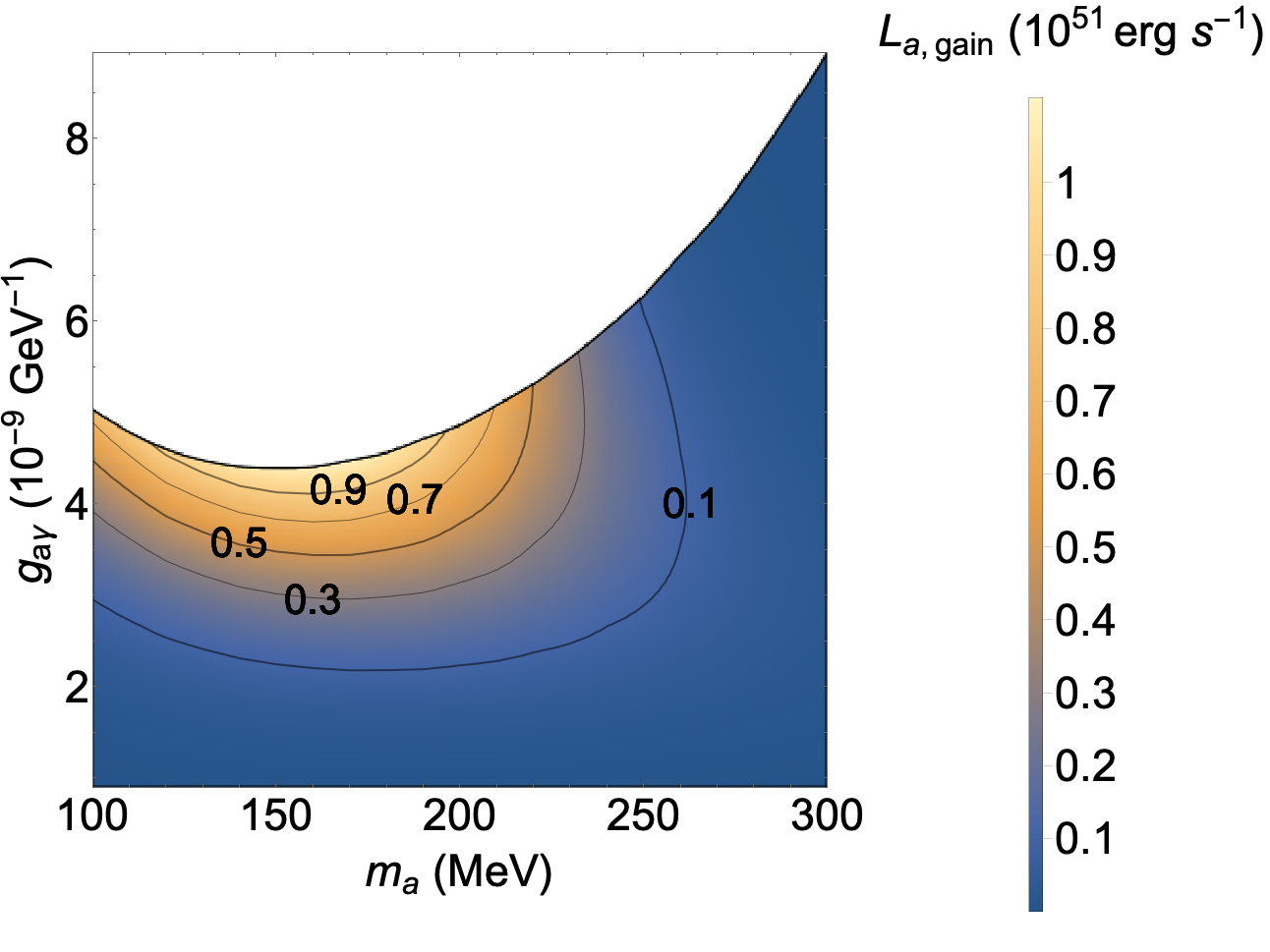} 
		\includegraphics[scale=0.185]{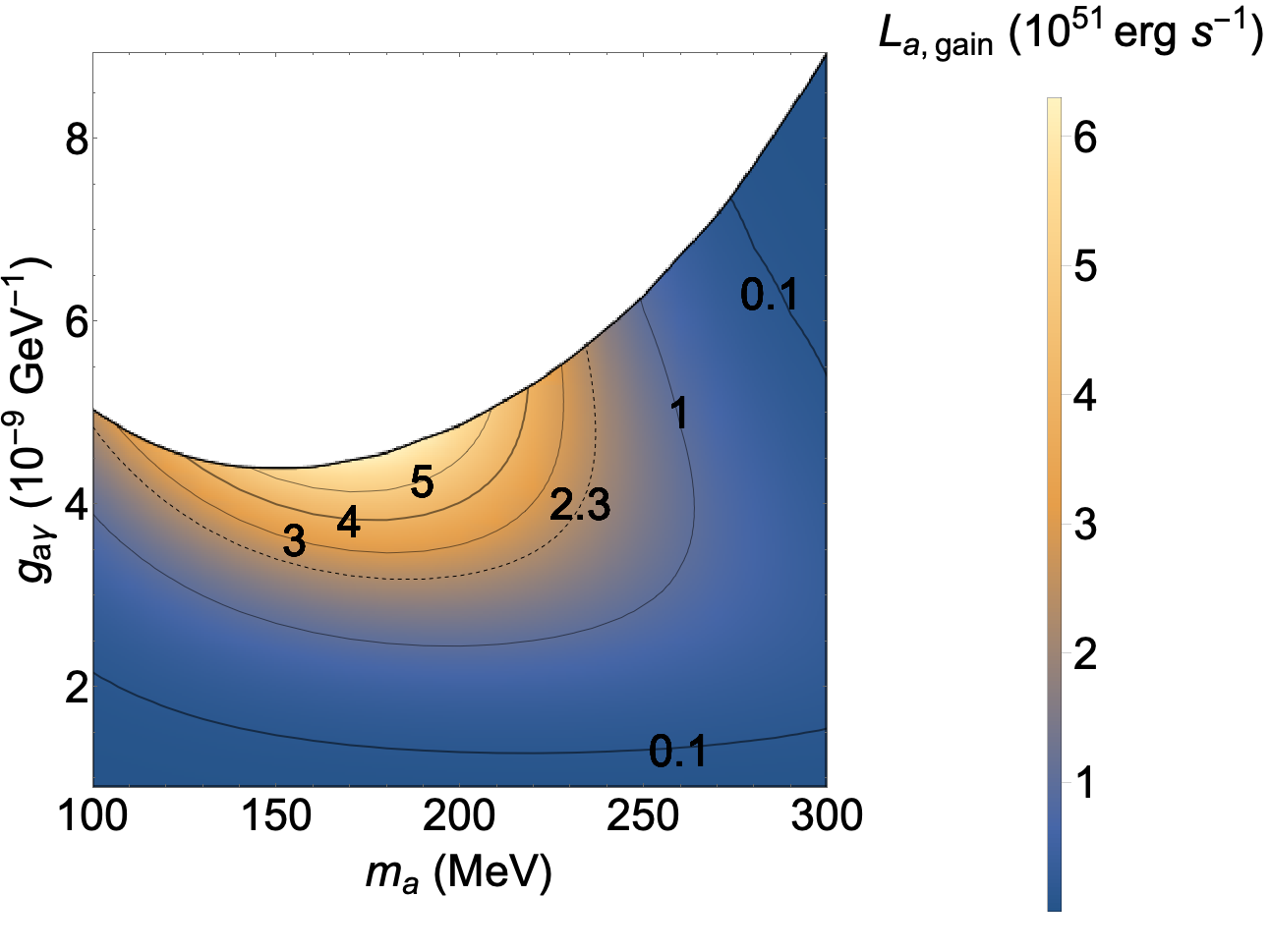}   
 		\includegraphics[scale=0.185]{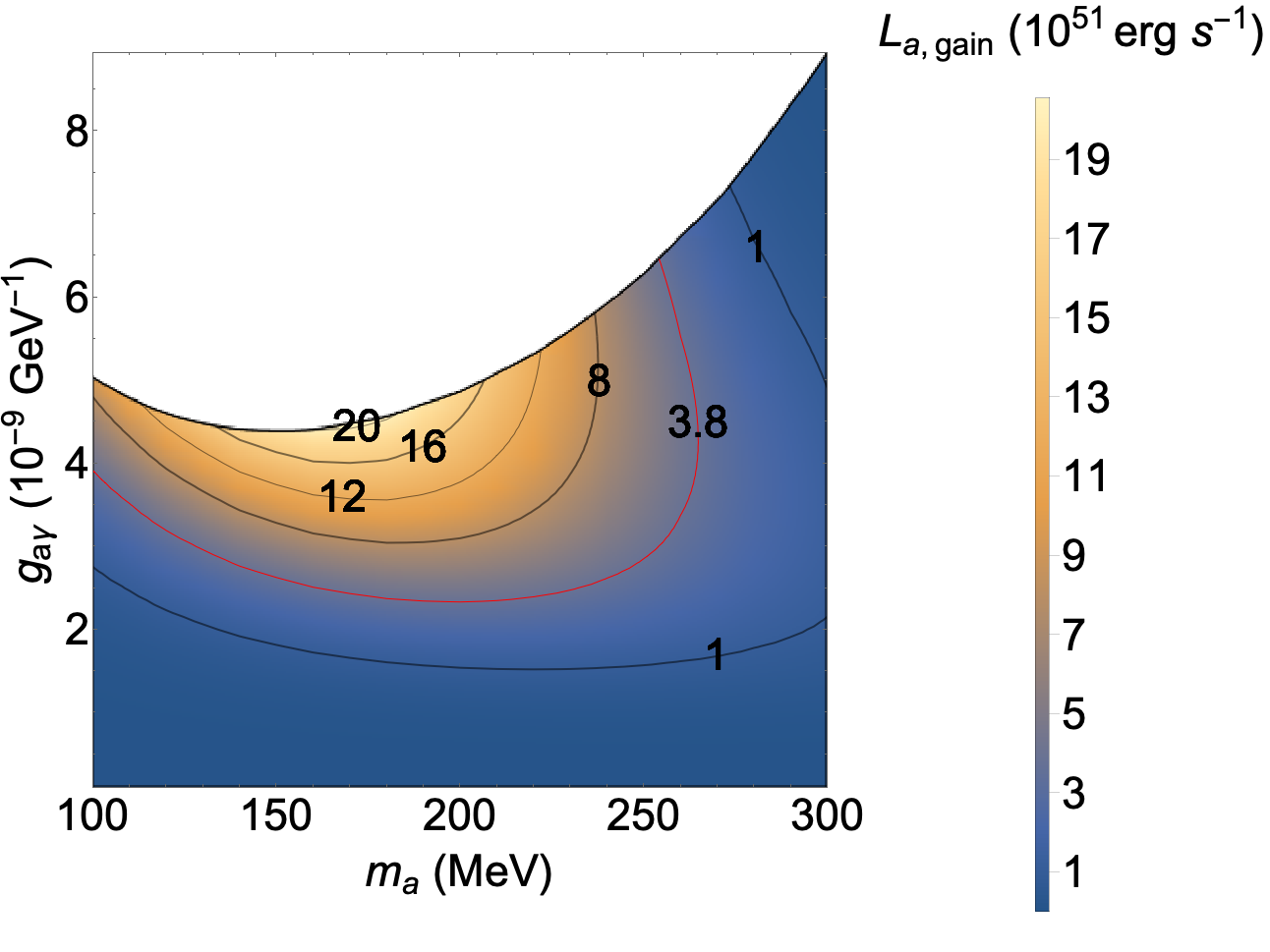} 
 	\caption{Isocontours in the $g_{a\gamma}-m_a$ plane of the ALP energy deposition rate in the gain layer at $t_\text{pb}=0.1$~s (upper
	panel), $t_\text{pb}=0.2$~s (middle panel),  $t_\text{pb}=0.3$~s (lower panel).
 {The dashed contour in the middle panel corresponds to the one tenth of $L_{\nu,\text{gain}}$ at $t_\text{pb}=0.2$~s, while the red contour in the lower panel corresponds to  $L_{\nu,\text{gain}}\approx3.8\times10^{51}$~erg~s$^{-1}$ at $t_\text{pb}=0.3$~s. In all the panels, the white region is excluded by the energy-loss argument.}}
 	\label{lumgain0103}
 \end{figure} 
 %%%%%%%%%%%%%%%%%%%%%%% 
In Fig.~\ref{lumgain0103} we show the contour plot in the $g_{a\gamma}-m_a$ plane of the ALP energy deposition rate in the gain layer
at different post-bounce times. 
 At $t_\text{pb}=0.1$~s (upper panel) the ALP energy deposition rate results to be $L_{a,\,\text{gain}}\sim O(10^{50}~\text{erg s}^{-1})$, two orders of magnitude smaller than the neutrino contribution at the same time.  At $t_\text{pb}=0.2$~s (middle panel) the ALP energy deposition rate in the gain layer is not negligible with respect to the neutrino one $L_{\nu,\,\text{gain}}\approx 2.3\times 10^{52}$ erg s$^{-1}$ (the dashed black contour {in the middle panel of Fig.~\ref{lumgain0103}} corresponds to one tenth of $L_{\nu,\,\text{gain}}$). Therefore, ALPs could help the trigger of the explosion for a range of parameters not excluded by the energy-loss argument, namely in the mass range $150$~MeV$\lesssim m_a\lesssim 220$~MeV and for values of the coupling constant $g_{a\gamma}\gtrsim 4\times 10^{-10}$~GeV$^{-1}$. After the explosion is triggered, the axion heating rate continues to increase and becomes much larger than the neutrino one since this latter rapidly decreases. In particular, at $t_\text{pb}=0.3$~s (lower panel) $L_{\nu,\,\text{gain}}\approx4\times10^{51}$~erg s$^{-1}$, while $L_{a,\,\text{gain}}\sim O(10^{52}~\text{erg s}^{-1})$.
 %%%%%%%%%%%%%
\begin{figure}[t!]
	\centering
		\includegraphics[scale=0.4]{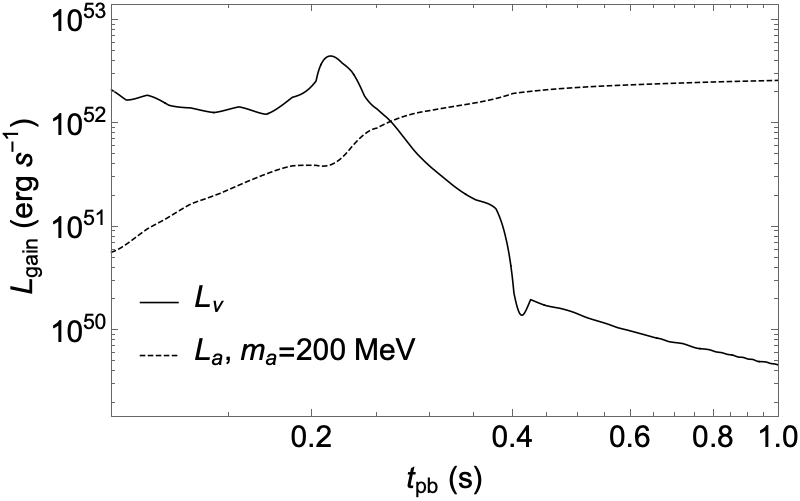} \\
		\vspace{8 pt}
		\includegraphics[scale=0.4]{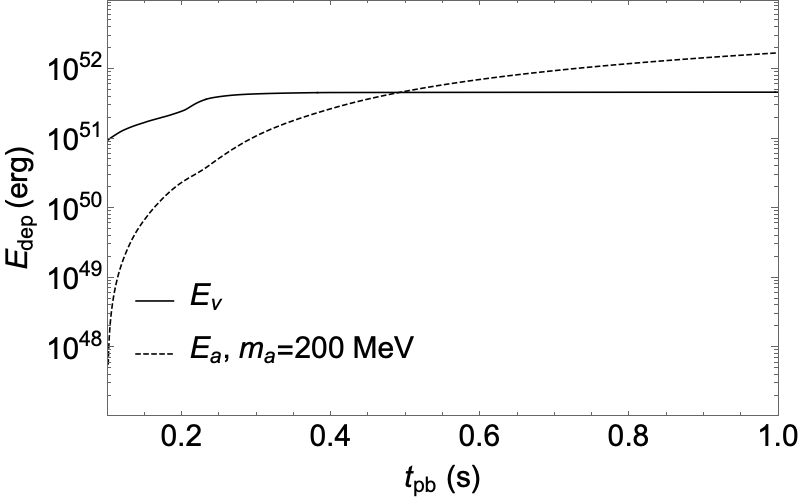} 
	\caption{The time evolution of the energy deposition rate (upper panel) and of the energy deposited (lower panel) in the gain layer by neutrinos (continuous black curves) and by ALPs (dashed black curves) with mass $m_a=200$~MeV and  coupling constant $g_{a\gamma}=4\times10^{-10}$~GeV$^{-1}$.}
	\label{lgaincfr}
\end{figure} 
%%%%%%%%%%%%%%%%%%%%%%

In Fig.~\ref{lgaincfr} we show the time evolution of the energy deposition rate (upper panel) and of the energy deposited by ALPs in the gain layer (lower panel) for a fixed value of the coupling constant $g_{a\gamma}=4\times10^{-10}$ GeV$^{-1}$ and a representative value of the axion mass $m_a=200$ MeV (dashed black curves). At early times ($t_\text{pb}<0.2$~s) $L_{a,\,\text{gain}}\ll L_{\nu,\,\text{gain}}$, while the ALP energy deposition rate becomes larger than $L_{\nu,\text{gain}}$ at times $t_{\text{pb}}\gtrsim 2.5$~s. Similarly, the energy deposited by ALPs in the gain layer is negligible at times $t_\text{pb}\lesssim0.2$~s since it is more than an order of magnitude smaller than the one deposited by neutrinos at the same time. However, the former constantly increases and becomes greater than $5\times10^{51}$~erg (the saturation value of the deposited neutrino energy) at times $t_\text{pb}\gtrsim 0.5$~s. \newline 
In particular, at $t_{\text{pb}}\approx0.3$~s, $E_{a,\,\text{dep}}\sim O(10^{51}~\text{erg})$, the mass of the gain layer is $M\sim O(0.1~\text{M}_\odot)$ and the temperature in this region is $T\approx 2$~MeV. Therefore, by Eq.~(\ref{deltas}) one has $\Delta s\approx$ a few units of Boltzmann's constant per baryon, an increase sufficient to partially melt nuclei and help the SN explosion. \newline
One may conclude that for axion couplings below the energy-loss bound, one can still have a 
non-negligible axion heating for the ALP parameters $150$~MeV$\lesssim m_a\lesssim 220$~MeV and $g_{a\gamma}\gtrsim4\times10^{-10}$~GeV$^{-1}$, which would help the SN explosion. In principle, one could convert the ALP deposited energy into an explosion energy
 $E_\text{expl}$~\cite{Fischer:2009af} and constrain the ALP parameter space by imposing
  $E_{\text{expl}}\lesssim3\times10^{51}$~erg~\cite{Sung:2019xie}. However, only a simulation including ALPs could provide reliable results and strengthen the validity of the hints obtained (see e.g.~\cite{Schramm:1981mk} for radiatively decaying standard axions or~\cite{Rembiasz:2018lok} for decaying sterile neutrinos).

%%%%%%%%%%%%%%
\section{Conclusions}
\label{sec:concl}
%%%%%%%%%%%%%%

In this work we used the state-of-the-art SN one-dimensional simulations to revise and update the mechanism 
of the emission from the SN core of heavy axion-like particles (ALPs), with masses of the order 1-100 MeV 
interacting with photons. 
In particular, we added the contribution of the photon coalescence process, neglected in previous studies, which dominates at masses $m_{a}\gtrsim100$~MeV and allows one to improve the previous constraints on the ALP parameter space. 
The detailed analysis of the SN simulations, which provides the time evolution of the characteristic radii of the SN atmosphere, allows one to consistently apply a recently proposed method (dubbed the ``modified luminosity criterion'') to explore the ALP parameter space, firstly presented in \cite{Chang:2016ntp} and based on the assumption that only the energy which cannot be efficiently reprocessed by the neutrino fluid is relevant to constraint the ALP parameters. Through this method, in the free-streaming regime we confirm the previous bounds in the low-mass limit [$m_a\lesssim 10~\text{MeV}$] since the photon coalescence contribution is suppressed as $m_a^{4}$, while constraints are sensitively improved at large masses. In particular, the inclusion of the photon coalescence allows us to strengthen the previous bounds in this regime by more than an order of magnitude for masses $m_a\gtrsim200$~MeV. On the other hand, in the trapping regime values of the coupling $g_{a\gamma}\lesssim 1.4\times10^{-5}$~GeV$^{-1}$ are excluded in the small mass limit, while the bound is relaxed for larger masses. In Fig.~\ref{overviewplot} we show the updated constraints on heavy ALPs. As discussed in Appendix~\ref{app:trapmethods}, in the trapping regime the modified luminosity criterion is more reliable than the constraining criterion based on ALP opacity used in prior works. However the uncertainties related to the SN model and the difficulties to deal with trapped ALPs suggest the importance to develop a SN simulation with a self-consistent inclusion of ALPs to have an even more reliable bound on the ALP parameter space.
%%%%%%%%%%%%%%%%%%
\begin{figure}[t!]
	\centering
	\includegraphics[scale=1]{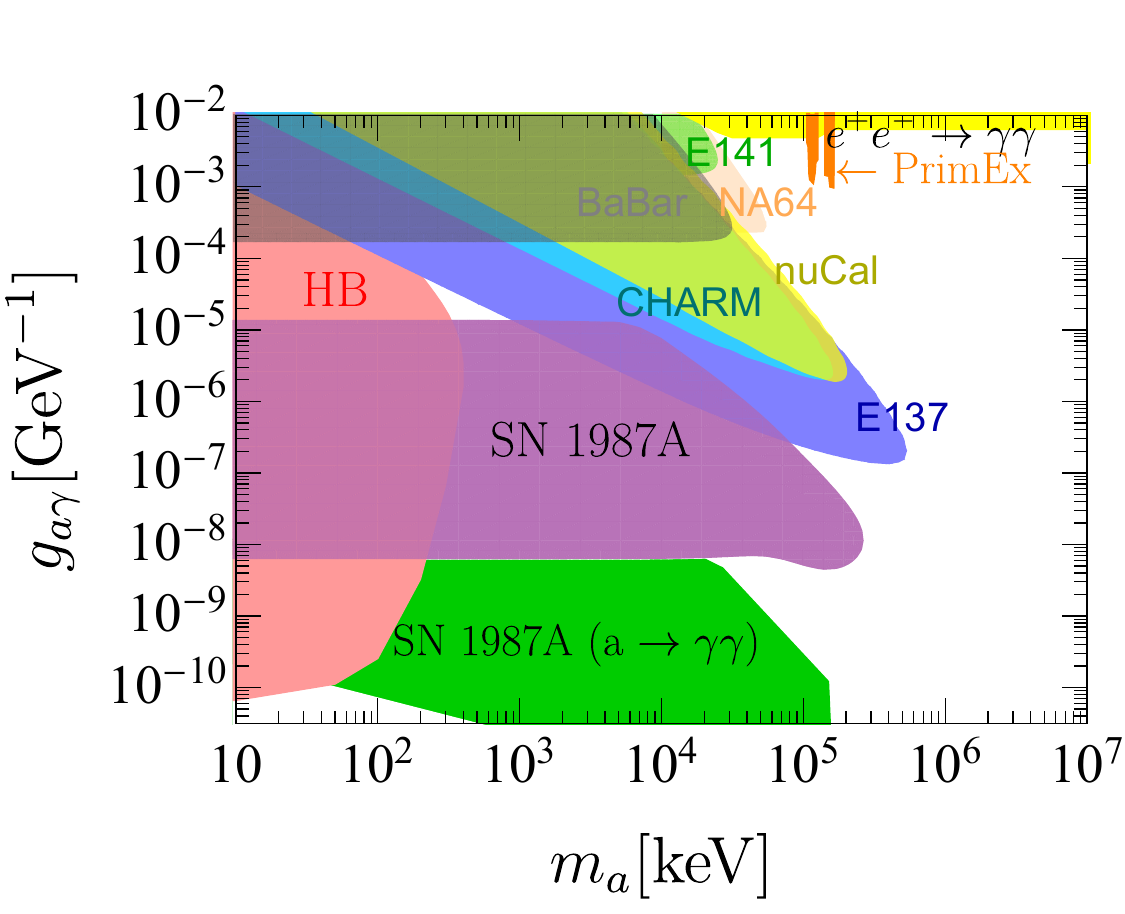} 
	\caption{Overview of the heavy ALP parameter space in the plane $g_{a\gamma}$ vs $m_a$. The purple-filled region labelled ``SN 1987A” represents our new exclusion result. The HB bound \cite{Carenza:2020zil}, the constraint from visible decays of ALPs produced in SN 1987A \cite{Jaeckel:2017tud} and the experimental limits, compiled from Refs.~\cite{Dolan:2017osp,Dobrich:2019dxc,Carenza:2020zil,Banerjee:2020fue}, are also shown.}
	\label{overviewplot}
\end{figure} 
%%%%%%%%%%%%%%%%

The early time evolution of the SN simulation allows one to assess the possible impact of the ALP production to trigger the SN explosion. In particular, following an original approach through which the neutrino heating rates and the ALP ones are compared, we endow a region of the parameter space with masses $m_a\approx 200$~MeV and coupling constant $g_{a\gamma}\gtrsim 4\times 10^{-9}$~GeV$^{-1}$ for which the ALPs decaying into photons would provide an efficient energy deposition behind the shock which could help the SN explosion. It is quite intriguing that ALPs could be identified as the yet-missing piece of the puzzle to boost the supernova explosion energy in present neutrino-driven explosion models. A further exploration of this aspect motivates the involvement of multi-dimensional supernova models. 
%%%%%%%%%

%%%%%%%%%%%%%%%%%%%%%%%%%%%%%%%%%%%%%%%%%%%%%%%%%%%%%%%%%%%%%%%%%%%%%%
\section*{Acknowledgments}
%%%%%%%%%%%%%%%%%%%%%%%%%%%%%%%%%%%%%%%%%%%%%%%%%%%%%%%%%%%%%%%%%%%%%%

We thank the anonymous referee for the valuable comments on our manuscript.
The work of P.C. and 
A.M. is partially supported by the Italian Istituto Nazionale di Fisica Nucleare (INFN) through the ``Theoretical Astroparticle Physics'' project
and by the research grant number 2017W4HA7S
``NAT-NET: Neutrino and Astroparticle Theory Network'' under the program
PRIN 2017 funded by the Italian Ministero dell'Universit\`a e della
Ricerca (MUR).
T.~F. acknowledges support from the Polish National Science Center (NCN) under Grant No. 2016/23/B/ST2/00720 and No. 2019/33/B/ST9/03059. The supernova simulations are performed at the Wroclaw Center for Scientific Computing and Networking (WCSS) in Wroclaw (Poland).

\appendix 

\section{Trapping regime: the opacity criterion}
\label{app:trapmethods}
A different approach to constrain the ALP parameter space in the trapping regime is based on the assumption that trapped ALPs may contribute significantly to the energy transport in the star, i.e. they would remove energy from one region of the PNS and deposit it at an approximate distance of one mfp, modifying the SN evolution. In this context, one defines the ALP Rosseland mfp as $\lambda_a=(\kappa_a\rho)^{-1}$ in terms of the Rosseland opacity \cite{Raffelt:1996wa}
%%%%%%%%%%%%%%%%%%%%%
\begin{equation}
\kappa_a^{-1}=\dfrac{\int_{m_a}^{\infty}\,\kappa_E^{-1}\,\beta_E\,\partial_TB_E\,dE}{\int_{m_a}^{\infty}\,\beta_E\,\partial_TB_E\,dE}\,,
\label{rosseland}
\end{equation}
%%%%%%%%%%%%%%%%%%
where
%%%%%%%%%%%%
\begin{equation}
B_E=\dfrac{1}{2\pi^2}\dfrac{E^2(E^2-m_a^2)^{1/2}}{e^{E/T}-1}\,
\end{equation}
%%%%%%%%%%%%%%%%%%%%%
is the ALP thermal spectrum and the ALP opacity $\kappa_E$ is evaluated
by considering in Eq.~(\ref{rosseland}) the contributions of  the inverse Primakoff effect $a+p\rightarrow p+\gamma$ and the decay process $a\rightarrow \gamma\gamma$
%%%%%%%%%%%%%%%%%%
\begin{equation}	\kappa_E=\kappa_{a\rightarrow\gamma}+\kappa_{a\rightarrow\gamma\gamma}\,.
\label{Ke}
\end{equation}
%%%%%%%%%%%%%%%%%%%
In particular, the decay contribution $\kappa_{a\rightarrow\gamma\gamma}$ results to be  $\kappa_{a\rightarrow\gamma\gamma}\rho= \lambda_{a\rightarrow\gamma\gamma}^{-1}$, where $\lambda_{a\rightarrow\gamma\gamma}$ is the decay mfp in Eq.~(\ref{mfpdec}), while the inverse Primakoff contribution $\kappa_{a\rightarrow\gamma}$ is $\kappa_{a\rightarrow\gamma}\rho= \lambda_{a\rightarrow\gamma}^{-1}$, with $\lambda_{a\rightarrow\gamma}$ the inverse Primakoff mfp in Eq.~(\ref{mfpprim}).

When the ALP mass is less than few MeV, the inverse Primakoff process dominates over the decay, which has a strong dependence on the axion mass ($m_a^3$) and is forbidden for $m_a < 2\omega_{\rm pl}$. 
Since the Primakoff opacity is strictly dependent on the matter density $\rho$, in the low-mass limit ALPs are trapped in the inner SN core and they are emitted from an ``axion-sphere'', the analogous of the ``neutrino-sphere'', with a radius $R_a$ determined by the condition 
%%%%%%%%%%%%%%%%%%%
\begin{equation}
\tau_a(R_a)=\int_{R_a}^\infty \kappa_{a}\,\rho\, dr=\dfrac{2}{3}\,,
\label{defra}
\end{equation}
%%%%%%%%%%%%%%%%%%%%%%
where $\kappa_a$ is the Rosseland mean opacity defined in Eq.~(\ref{rosseland}). 
Therefore, in this approach, for masses $m_a\lesssim O(1~\text{MeV})$, 
trapped ALPs are expected to have a black-body emission with a luminosity $L_a\propto R_a^2\,T^4(R_a)$. 
The bound on the coupling $g_{a\gamma}$ is obtained by imposing~\cite{Raffelt:1987yt}
%%%%%%%%%%%%%%%%
\begin{equation}
L_a\lesssim L_\nu\,.
\label{l}
\end{equation}
%%%%%%%%%%%%%%%%%%%
This condition excludes the values of the photon-axion coupling $g_{a\gamma}\lesssim 7.7\times 10^{-6}$~GeV$^{-1}$,
 in agreement with Ref.~\cite{Dolan:2017osp}.

\begin{figure}[t!!]
 	\centering
 	\includegraphics[scale=0.38]{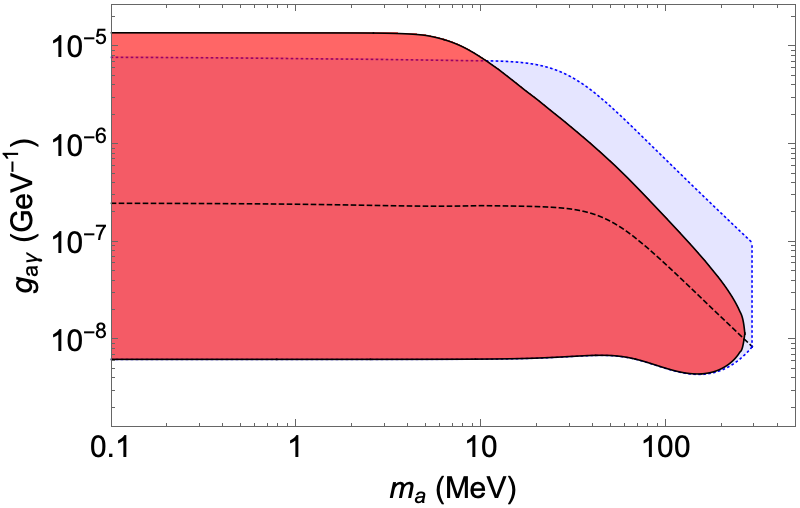}   \\
 	\caption{Exclusion plot in the plane $m_a$--$g_{a\gamma}$. The shaded red area is excluded by the modified luminosity criterion, while the light blue area is excluded by splitting the calculation in two different regimes (details in the text). The dashed black line represents the border between the free-streaming and trapping regimes.}
 	\label{cfrbound}
 \end{figure}

\begin{figure}[t!!]
 	\centering
 	\includegraphics[scale=0.38]{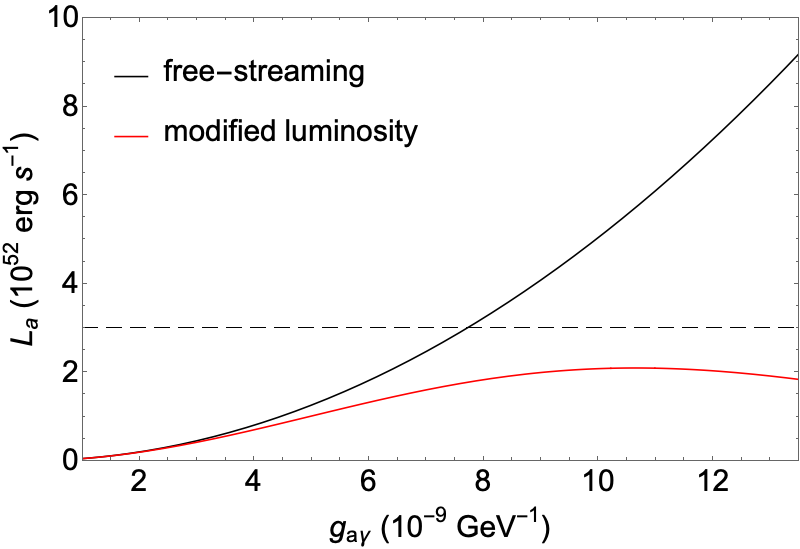}   \\
 	\caption{The ALP luminosity in the free-streaming approximation Eq.~(\ref{luma}) (black line) and the modified luminosity Eq.~(\ref{lumax}) (red line) as a function of the coupling constant $g_{a\gamma}$ for an ALP with mass $m_a=280$ MeV. The dashed horizontal line represent the critical value $L_\nu=3\times10^{52}$ erg s$^{-1}$.}
 	\label{lum280}
 \end{figure}

For heavier ALPs, $m_a\gtrsim 10$~MeV, the decay process becomes dominant. This implies that at these masses the ``axion-sphere'' cannot be well defined because the integral in Eq.~(\ref{defra}) always diverges since the decay mfp tends to a constant value in vacuum.
%Therefore ALPs may decay before leaving the SN core, 
%contributing to the energy transfer. 
In this context, ALPs may decay before leaving the SN core, contributing to the energy transfer. In order to constrain $g_{a\gamma}$ one should impose at the neutrino-sphere $R_{\nu}$ that
\begin{equation}
\kappa_a\gtrsim\kappa_\nu\,,
\label{k}
\end{equation}
 where $\kappa_\nu$ is the neutrino opacity, given by Eqs.~(\ref{opnu})--(\ref{keff}).
  The bound obtained following this approach has to be connected with the one for the low-mass case dominated by the Primakoff process, computed through Eq.~(\ref{l}). 
  The resulting constraint is constant in the low-mass limit $m_a\lesssim10$~MeV 
while for higher masses it decreases as $\sim m_a^{-3/2}$, as shown in Fig.~\ref{cfrbound}, where the shaded red region is excluded through the modified luminosity criterion, while the light blue region delimited by a dotted blue line represents the region excluded by combining the energy-loss argument in the free-streaming and the energy-transfer argument in the trapping regime. 
 %%%%%%%%%%%%%%%%%%%
 %%%%%%%%%%%%%%%%
A disadvantage of the approach described in this Appendix is that one has to split the calculation into two different regimes, while with the modified luminosity criterion the transition from the free-streaming to the trapping regime is smooth due to the presence of the optical depth factor. Therefore, in this case one needs a strategy to assess for which values of the ALP parameters the free-steaming approximation is good. In particular, the dashed black line in Fig.~\ref{cfrbound} represents the border between the free-streaming (lower $g_{a\gamma}$) 
and the trapping regime (higher $g_{a\gamma}$),
obtained by imposing the condition~\cite{Raffelt:1996wa}
 \begin{equation}
 \tau_a(R_p,R_\nu)=\int_{R_p}^{R_\nu} dr \kappa_a \rho = 1\,,
 \label{op1}
 \end{equation} 
 where $R_p$ is the mean radius at which the axions are produced
 \begin{equation}
R_p=\dfrac{\int dr \,r\, Q_a(r)}{\int dr\, Q_a(r)}\,.
\label{Rp}
 \end{equation}
 The condition in Eq.~(\ref{op1}) corresponds to the requirement that an ALP produced at $R_p$ 
 should emerge from the neutrino-sphere, with a survival probability $e^{-1}$. The dashed line intersects the ``modified luminosity bound'' at the transition between the two regimes at $m_a\approx266$ MeV, while it crosses the free-streaming bound at $m_a\approx290$~MeV. Therefore
 the energy-loss argument is surely no more reliable for higher masses, 
 which is the reason why the exclusion region has been cut with a vertical line at this value of the mass. However, the free-streaming approximation is not completely reliable for masses just below this limit too. As shown in Fig.~\ref{lum280}, for an ALP with mass $m_a=280$ MeV in the free-streaming approximation Eq.~(\ref{luma}) values of the coupling $g_{a\gamma}\gtrsim7.7\times10^{-9}$ GeV$^{-1}$ are excluded, while the modified luminosity Eq.~(\ref{lumax}) (which takes into account the axion decay, non-negligible for heavy ALPs) does not violate the bound for any value of the coupling constant.
 %%%%%%%%%%%%%%%%%%%
 %%%%%%%%%%%%%%%%
 
    %%%%%%%%%%%%%%%%%%%
 \begin{figure}[t!!]
 	\centering
 	\includegraphics[scale=0.38]{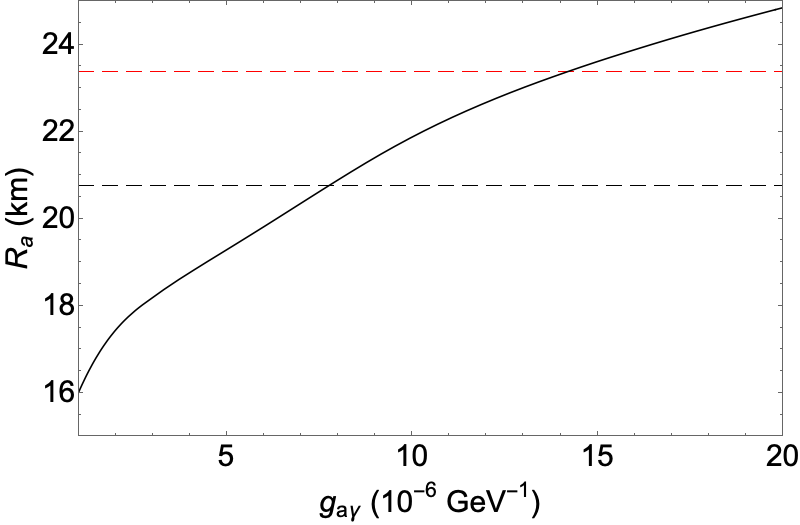}   \\
 	\vspace{10 pt}
 	\includegraphics[scale=0.37]{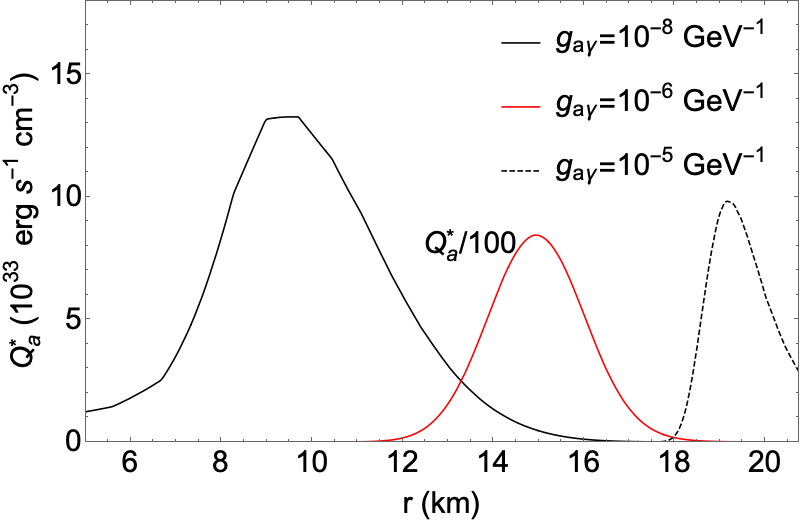}
 	\caption{\textit{Upper panel}: The axion-sphere radius $R_{a}$ as a function of the coupling constant $g_{a\gamma}$ for an ALP with mass $m_{a}=100$ keV. The dashed horizontal lines correspond to $R_{\nu}$ (black) and $R_{\rm{gain}}$ (red). \textit{Lower panel}: The integrand in Eq.~(\ref{lumax}) $Q_{a}^*\equiv Q_{a}(r)e^{-\tau_{a}^{*}(r,R_{\text{far}})}$ as a function of the radius for an ALP with mass $m_{a}=100$ keV and different values of $g_{a\gamma}$ as shown in legend. Notice that for $g_{a\gamma}=10^{-6}$ GeV$^{-1}$ (red curve) $Q_{a}^{*}$ has been divided by a factor 100 to be of the same order of magnitude of the other quantities in the plot.}
 	\label{funzg}
 \end{figure} 
 %%%%%%%%%%%%%%%
 While for the free-streaming the only problem is related to the validity limit of this regime, the situation for the trapping regime is more complex.
 In particular, in the low-mass limit ($m_a\lesssim10$~MeV) the modified luminosity criterion allows one to exclude values of the coupling larger than those excluded by evaluating the ALP emission at the axion-sphere radius in Eq.~(\ref{defra}). Indeed, the black-body bound ($g_{a\gamma}\lesssim7.7\times10^{-6}$ GeV$^{-1}$) ceases because the axion-sphere moves outside from the neutrino-sphere, as shown in the upper panel of Fig.~\ref{funzg}. On the other hand, for slightly larger values of the coupling constant, although $R_a>R_\nu$, the modified luminosity remains large enough to violate the bound. This suggests that the thermal emission underestimates the effective luminosity. In the lower panel of Fig.~\ref{funzg} we show the integrand in Eq.~(\ref{lumax}) $Q_a^*\equiv Q_a(r)e^{-\tau_a^*(r,R_\text{far})}$ as a function of the radius for increasing values of $g_{a\gamma}$. As the coupling constant increases, the integrand becomes narrower but it does not become a delta function, therefore the black-body emission underestimates the real luminosity since the production region is quite larger than a fixed radius typical of a thermal emission.
 Additionally, in the region of large mass and high coupling the bound obtained by comparing the opacities is larger than the one obtained through the modified luminosity criterion. Indeed, for the latter criterion the axions have to go out from the gain radius to contribute to the energy loss. Therefore, the reason for the discrepancy in this region is that there are values of the ALP parameters for which the axion mfp is longer than that of the neutrino, hence violates the opacity comparison, but it is still shorter than $R_{\rm{gain}}-R_\nu$, thus the axions do not go far enough to modify the neutrino signal. \newline
 Since a self-consistent inclusion of ALPs in a SN simulation is necessary to show that the SN evolution actually changes if the axion mfp is larger than the neutrino one, without a SN simulation which takes into account the axions feedback the ``modified luminosity criterion'' is surely the most reliable strategy to constrain the ALP parameter space.
 
 %%%%%%%%%%%%
% \begin{figure}[t!!]
% 	\centering
% 	\includegraphics[scale=0.31]{lumfunzg}   \\
% 	\caption{The integrand in Eq.~(\ref{lumax}) $Q_a^*\equiv Q_a(r)e^{-\tau_a^*(r,R_\text{far})}$ as a function of the radius for different values of $g_{a\gamma}$ as shown in legend. Notice that for $g_{a\gamma}=10^{-6}$ GeV$^{-1}$ (red curve) $Q_a^*$ is divided by a factor 100 to be of the same order of magnitude of the other quantities in the plot.}
 %	\label{lumfunzg}
 %\end{figure} 
 %%%%%%%%%%%%%%%%

\section{Effect of the progenitor mass on the bound}
\label{app:uncertainties}
%%%%%%%%%%%%%%%%%%
    \begin{figure}[t!]
	\centering
		\includegraphics[scale=0.35]{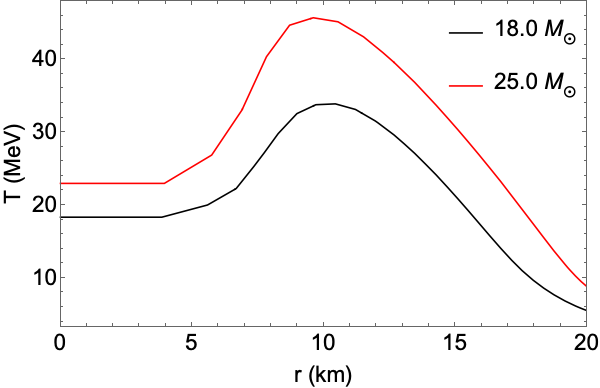} 
		\includegraphics[scale=0.35]{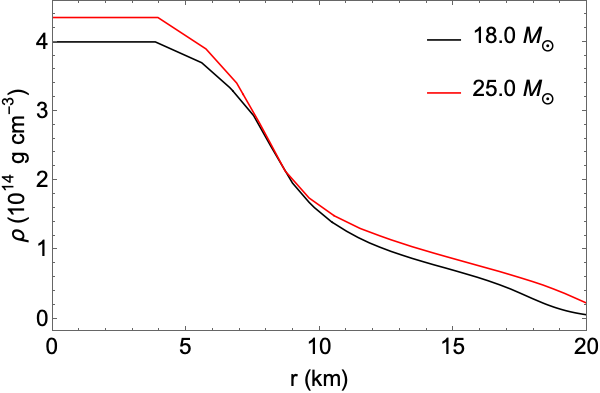}  \\
		\vspace{10 pt}
		\includegraphics[scale=0.35]{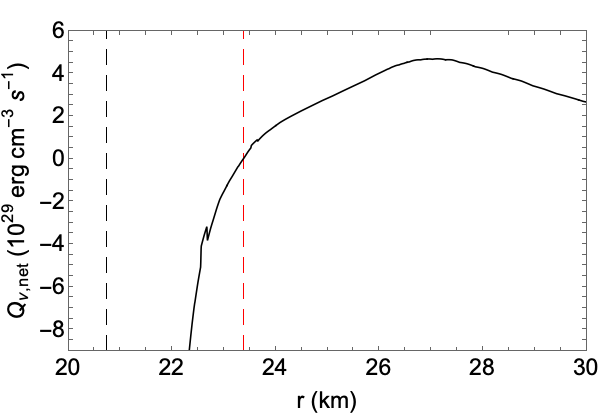} 		\includegraphics[scale=0.35]{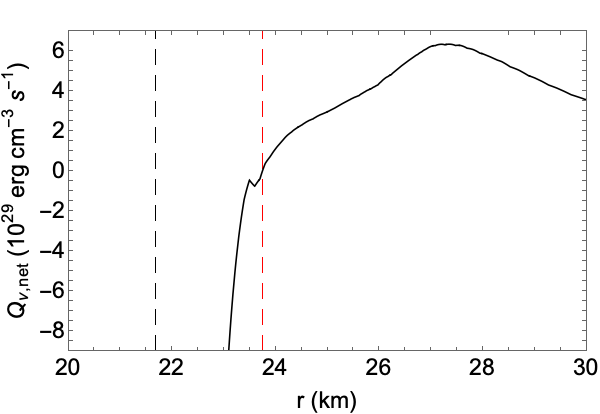} 
	\caption{\textit{Upper Left Panel}: The radial profile of the temperature $T$ at $t_{\rm{pb}}=1$~s for the 18 $M_\odot$ (black line) and the 25 $M_{\odot}$ (red line) model. \textit{Upper Right Panel}: The radial profile of the matter density $\rho$ at $t_{\rm{pb}}=1$~s for the 18 $M_\odot$ (black line) and the 25 $M_\odot$ (red line) model. \textit{Lower Left Panel}: The neutrino net heating rate $Q_{\nu,\rm{net}}$ for the 18 $M_\odot$ model at $t_{\rm{pb}}=1$~s. \textit{Lower Right Panel}: The neutrino net heating rate $Q_{\nu,\rm{net}}$ for the 25 $M_\odot$ model at $t_{\rm{pb}}=1$~s. In the lower panels, the vertical dashed lines correspond to the neutrino-sphere radius (black) and the gain radius (red).}
	\label{temp25}
\end{figure}
\begin{figure}[t!]
	\centering
		\includegraphics[scale=0.4]{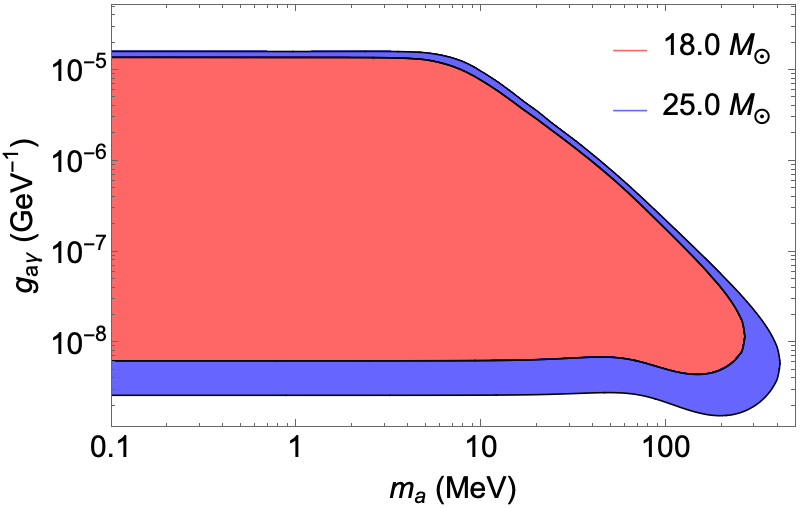} 
	\caption{The ALP exclusion plot in the plane $m_a-g_{a\gamma}$ obtained through the modified luminosity criterion. The red region corresponds to the exclusion plot for a 18 $M_\odot$ case, while the blue one refers to 25 $M_\odot$.}
	\label{bound25}
\end{figure}

 
%%%%%%%%%%%%%%%%
In order to assess how the SN 1987A bound is affected by the model used, in this Section we re-evaluate it choosing a different progenitor mass. In particular, we consider a SN model with the same equations of state but a progenitor mass of $25$ $M_\odot$, larger than the one of our reference model (18 $M_\odot$). In Fig.~\ref{temp25} the radial profiles of the temperature $T$ (upper left panel) and the matter density $\rho$ (upper right panel) at $t_{\rm{pb}}=1$ s for the 18 $M_\odot$ (black curves) and 25 $M_\odot$ (red curves) models are compared. In the ALP production region ($5$~km$\lesssim r \lesssim15$~km) the temperature is about 10 MeV larger in the 25 $M_\odot$ model, while the matter density does not change very much. In order to compute the ``modified luminosity'' in Eq. (\ref{lumnuova}) at $t_{\rm{pb}}=1$~s, we assume the core radius $R_c=10$~km, while by using the same approach described in Sec.~\ref{sec:radii} the neutrino-sphere radius and the gain radius result to be respectively $R_\nu=21.7$~km and $R_{\rm{gain}}=23.7$~km for the $25$ $M_\odot$ model, slightly larger than the corresponding quantities in the reference model, as shown in the lower panels of Fig.~\ref{temp25}. \newline
In Fig.~\ref{bound25} we show how the bound changes with the SN model. It is apparent that through the 25 $M_{\odot}$ model the exclusion region is larger than the one excluded with the 18 $M_{\odot}$ model. In particular, the discrepancy in the free-streaming regime is more substantial than the one in the trapping regime. Indeed, the core temperature of the star increases substantially when moving to larger progenitor masses. Since the axion production rate is a steep function of the temperature, in the free-streaming regime the energy loss in ALPs is much larger and thus the bound is stronger. On the other hand, in the trapping regime the bound does not change very much. In this regime the bound depends on the difference $R_{\rm{gain}}-R_\nu$ which slightly increases with the progenitor mass. Therefore the bound in the trapping regime is slightly stronger for larger progenitor mass.
%, but the discrepancy is not so large as in the free-streaming case. 
Naively, the same results can be obtained also with the ``black-body approximation''. 
Since axions thermalize in the trapping regime, the luminosity depends only on the temperature of the surface emission and a lot of information about the star is lost, therefore the bound is not very sensitive to the SN model.\newline

%%%%%%%%%%%%%%%%%%

%%%%%%%%%%%%%%%%%%%%%%%%%%%%%%%%%%%%%%%%%%%%%%%%%%%%%%%%%%%%%%%%%%%%%%
%\section*{References}
%%%%%%%%%%%%%%%%%%%%%%%%%%%%%%%%%%%%%%%%%%%%%%%%%%%%%%%%%%%%%%%%%%%%%%

\end{document}